\documentclass[a4paper,12pt]{article}
\pdfoutput=1

\usepackage{amsmath, amssymb, authblk, a4wide, bm, cancel, color, graphicx, multirow, subfig, enumerate, appendix, cite, hhline, hyperref, bbold, verbatim, slashed, float, tikz}

\usetikzlibrary{positioning,arrows,patterns}
\usetikzlibrary{decorations.pathmorphing}
\usetikzlibrary{decorations.markings}
\usetikzlibrary{calc}
\tikzset{
    photon/.style={decorate, decoration={snake}, draw=black, thick},
    fermionnoarrow/.style={draw=black, postaction={decorate}, thick},
    scalar/.style={draw=black, postaction={decorate}, thick, dashed},
    fermion/.style={draw=black, postaction={decorate},decoration={markings,mark=at position .55 with {\arrow{>}}}, thick},
    gluon/.style={decorate, draw=black, decoration={coil,amplitude=4pt, segment length=5pt}, thick},
    vertex/.style={draw,shape=circle,fill=black,minimum size=3pt,inner sep=0pt} 
}

\def\be{\begin{equation}}
\def\ee{\end{equation}}
\def\ba{\begin{eqnarray}}
\def\ea{\end{eqnarray}}

\def\p0{{\phantom{+}0}}

\def\uno{\mbox{1 \kern-.59em {\rm l}}}
\newcommand{\ampA}{\mathcal{A}}
\newcommand{\lag}{\mathcal{L}}
\newcommand{\ampM}{\mathcal{M}}

\numberwithin{equation}{section}

\begin{document}

\title{
\vspace{-3cm}
\rightline{\sc\small UMN-TH-3506/15}
\vspace{2cm}
\Large{\textbf{Long-Lived, Colour-Triplet Scalars from Unnaturalness}}}
\author[1]{\small{\bf James Barnard}\thanks{\texttt{james.barnard@unimelb.edu.au}}}
\author[1]{\small{\bf Peter Cox}\thanks{\texttt{pcox@physics.unimelb.edu.au}}}
\author[2]{\small{\bf Tony Gherghetta}\thanks{\texttt{tgher@umn.edu}}}
\author[1,3]{\small{\bf Andrew Spray}\thanks{\texttt{a.spray.work@gmail.com}}}
\affil[1]{\footnotesize ARC Centre of Excellence for Particle Physics at the Terascale, School of Physics,
The University of Melbourne, Victoria 3010, Australia }
\affil[2]{\footnotesize School of Physics and Astronomy, University of Minnesota, Minneapolis, Minnesota 55455, USA}
\affil[3]{\footnotesize Center for Theoretical Physics of the Universe, Institute for Basic Science (IBS), Daejeon, 34051, Korea} 
\date{}
\maketitle

\vspace{-1cm}
\begin{abstract}
\baselineskip=15pt
\noindent
Long-lived, colour-triplet scalars are a generic prediction of unnatural, or split, composite Higgs models where the spontaneous global-symmetry breaking scale $f \gtrsim 10$~TeV\@ and an unbroken $SU(5)$ symmetry is preserved. Since the triplet scalars are pseudo Nambu-Goldstone bosons they are split from the much heavier composite-sector resonances and are the lightest exotic, coloured states. This makes them ideal to search for at colliders. Due to discrete symmetries the triplet scalar decays via a dimension-six term and given the large suppression scale $f$ is often metastable. We show that existing searches for collider-stable R-hadrons from Run-I at the LHC forbid a triplet scalar mass below 845 GeV, whereas with 300~fb$^{-1}$ at 13~TeV triplet scalar masses up to 1.4~TeV can be discovered. For shorter lifetimes displaced-vertex searches provide a discovery reach of up to 1.8~TeV\@. In addition we present exclusion and discovery reaches of future hadron colliders as well as indirect limits that arise from modifications of the Higgs couplings.
\end{abstract}

\newpage
\tableofcontents

\section{Introduction}

The discovery of a light Higgs boson and the conspicuous absence of new states beyond the Standard Model at Run-I of the Large Hadron Collider (LHC) suggests that the scale of new physics may well be beyond that suggested by naturalness arguments. Composite Higgs models (for a recent review see \cite{1506.01961}), which are typically motivated as a possible solution to the hierarchy problem, have therefore come under increased scrutiny as lower limits on resonance masses strain the boundaries imposed by naturalness. This tension is further exacerbated by precision electroweak and flavour constraints, both of which prefer a much larger value of the spontaneous global-symmetry breaking scale, $f$, than can be directly probed at the LHC.

A simple solution that can satisfy even the most stringent constraints (typically due to flavour) is to require that $f\gtrsim 10$~TeV\@. This leads to an unnatural, or split, composite Higgs model~\cite{Barnard:2014tla} in which the Higgs mass-squared is tuned to the order of $10^{-4}$ and the particle spectrum splits into light pseudo Nambu-Goldstone bosons and heavy composite-sector resonances. Despite their unnaturalness these models still preserve gauge coupling unification due to the presence of a composite right-handed top quark and an unbroken $SU(5)$ symmetry in the composite sector provided $f\lesssim 500$ TeV. An immediate consequence is that the low-energy spectrum always contains a colour-triplet, pseudo Nambu-Goldstone boson; the colour-triplet partner of the composite Higgs doublet. In addition discrete symmetries, which arise from proton stability, furnish these models with a singlet scalar dark matter candidate, $S$. In the minimal model, the same discrete symmetries imply that the colour-triplet scalar decays to quarks and a pair of singlet scalars via a dimension-six term in the low-energy, effective Lagrangian. Since this high-dimension term is suppressed by the large symmetry-breaking scale, $f\gtrsim10$~TeV, the triplet-scalar is often metastable. Long-lived, colour-triplet scalars therefore provide a unique way to test unnaturalness in composite Higgs models.
 
\begin{figure}[!t]
  \centering
  \includegraphics[width=0.6\textwidth]{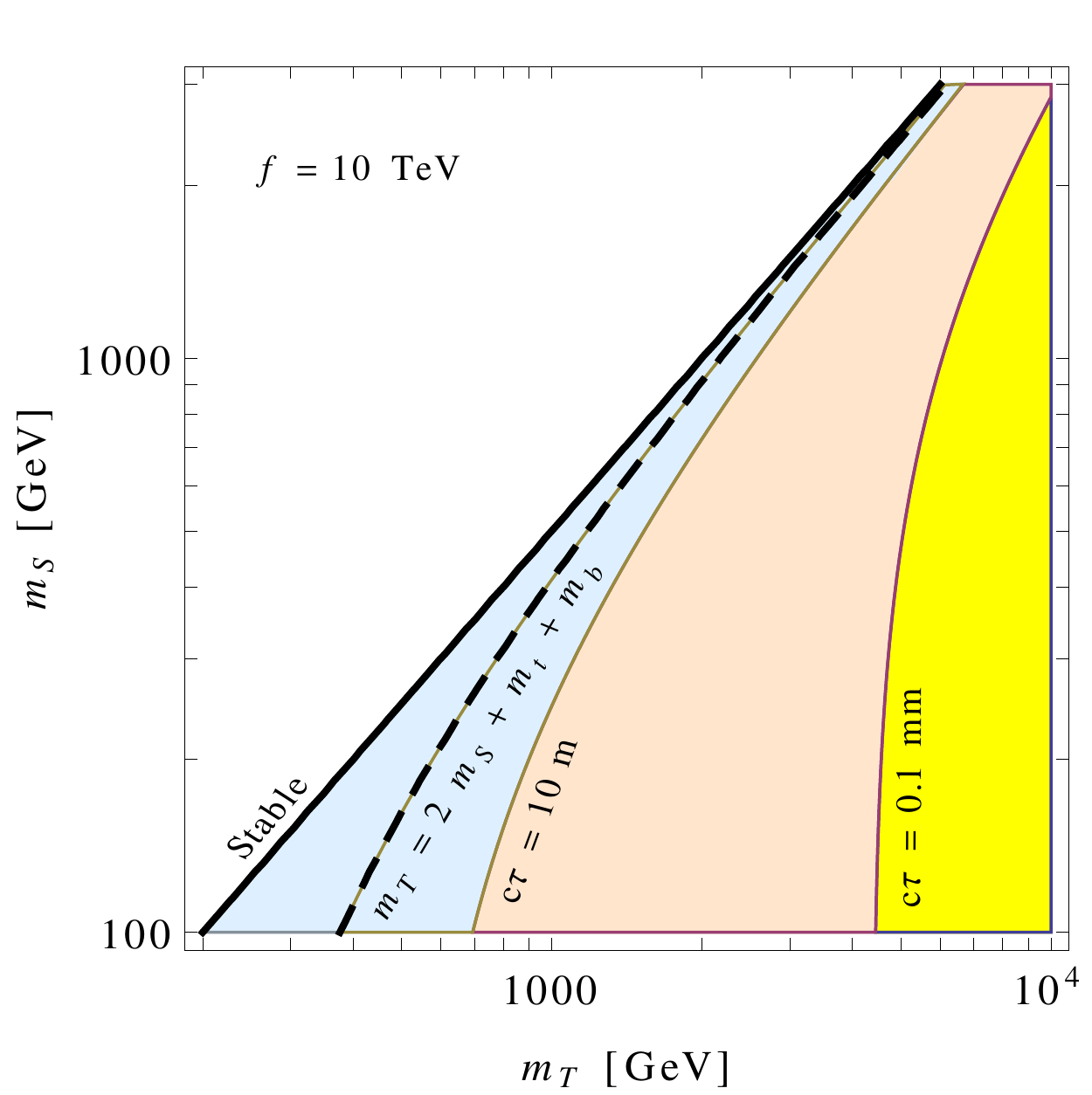}
  \caption{A schematic diagram of the possible types of decays as a function of the colour-triplet scalar mass $m_T$ and singlet scalar mass $m_S$. The three shaded regions from left to right correspond to decays that are collider stable, displaced and prompt, respectively. The dashed line represents the kinematic limit for the decay $T\rightarrow t^c b^c SS$ and the black solid line represents the limit when $m_T = 2 m_S$.}
  \label{fig:decays}
\end{figure}

Motivated by unnatural composite Higgs models we study the collider limits on long-lived, colour-triplet scalars and the prospects for detecting them at future colliders. The colour-triplet will be pair-produced via QCD processes and has the same quantum numbers as a right-handed scalar bottom quark. If long-lived, a colour-triplet will hadronize to form an R-hadron and can be detected in various ways depending on its decay length. The range of decay lengths as a function of the singlet mass $m_S$ and triplet mass $m_T$ is shown in Figure~\ref{fig:decays}. 

First, if the colour-triplet scalar is collider stable (i.e.\ decaying outside the detector), charged R-hadrons will leave a track in the inner detector and possibly the muon chamber. R-hadron searches at the LHC can then be used to place mass limits on the colour-triplet. Current limits from LHC Run-I results forbid a collider-stable colour-triplet with a mass below about 845~GeV\@. At Run-II similar searches will be performed and we show that with 300~fb$^{-1}$ of integrated luminosity triplet masses up to about 1.4~(1.5)~TeV can be discovered (excluded) for lifetimes corresponding to $c\tau\gtrsim10$~m. The discovery reach is significantly increased at a 100 TeV proton collider where discovery of a colour-triplet scalar with a mass up to 2-6~TeV, depending on its lifetime, will be possible, otherwise exclusion limits ranging from 2-7~TeV can be set. These limits depend only on the mass and width of the colour-triplet, therefore the results we obtain are quite general and can be applied to any other model predicting a similar, long-lived particle.

A second possibility is that the colour-triplet scalar decays within the detector (at radial distances greater than about 4~mm) and produces a displaced vertex (DV)  in the inner detector. The colour-triplet in the minimal model decays into a top quark, bottom quark and two singlet scalars, so the collider signature is predominantly jets from the quarks and missing energy from the singlets. This signal has previously been used to search for long-lived superparticles such as gluinos and squarks. While current results from displaced searches do not constrain the colour-triplet mass, these searches will become increasingly important at Run-II and beyond. With 300~fb$^{-1}$ at $\sqrt{s}=13$~TeV we find that colour-triplet masses up to 1.8~(1.9)~TeV can be discovered (excluded) for singlet masses below 450~GeV. In the future a 100~TeV collider would significantly improve the discovery reach, up to colour-triplet masses in the range 3-10~TeV depending on the singlet mass.

The final possibility is that the colour-triplet scalar decays promptly, dominantly producing jets and missing energy. These decays become relevant when the colour-triplet is heavier than about 4~TeV\@. For such heavy colour-triplets the production cross section at LHC energies is quite small and there will be too few events to detect them, even at the high-luminosity (HL) LHC\@. Instead, prompt decays could be searched for at a hypothetical 100 TeV proton collider. Using a similar search strategy to that used for gluinos we show that a future collider is potentially able to exclude colour-triplet masses in the range 4-7~TeV for singlet masses in the range 100-900~GeV.

Indirect limits on the colour-triplet scalar mass can be obtained by constraining modifications to the Higgs couplings. Using the current LHC results we find that colour-triplet masses are mostly constrained by the Higgs coupling to gluons to be in the range $m_T\gtrsim 100$~GeV. This limit will improve at the HL-LHC and ILC, although the most robust limits are inferior to the bound obtained from requiring that the triplet be heavier than twice the singlet scalar mass. The latter is constrained by direct detection experiments, with the current LUX bound giving $m_S\gtrsim150$~GeV and hence $m_T\gtrsim300$~GeV.

Previous studies of long-lived particles have primarily focused on supersymmetric models, motivated by the idea of split supersymmetry~\cite{Wells:2003tf, ArkaniHamed:2004fb,1210.0555,1212.6971} or simplified toy models with R-parity violation~\cite{1409.6729,1503.05923,1504.07293,1505.00784,1505.03479,1506.08206}. Our work is the first analysis of models based on the composite Higgs idea. It is a complete framework, incorporating gauge coupling unification, dark matter and an explanation for the fermion mass hierarchies, that represents an alternative to split supersymmetric models. Interestingly, unnatural (or split) composite Higgs models lead to similar decay signals albeit with different properties of the decaying particle and decay products. It will therefore be interesting to experimentally distinguish between these two ideas at future colliders.

The outline for the rest of this paper is as follows. In Section~\ref{sec:GF} we review the unnatural composite Higgs model and derive the decay width and corresponding decay length for the colour-triplet scalar. The limits from experimental searches are presented in Section~\ref{sec:ES}. We first discuss direct limits from R-hadron searches at the LHC and a future 100 TeV collider, followed by limits from displaced-vertex searches and limits from prompt decays. We then end with indirect limits that are obtained by studying modifications of the Higgs couplings. We summarise our results in Section~\ref{sec:CO}. Details of the four-body phase space integral are given in Appendix~\ref{app:phase space} and in Appendix~\ref{app:DV_validation} we compare the validity of our assumptions on the displaced-vertex search with the full experimental analysis.

\section{The Unnatural Composite Higgs Model\label{sec:GF}}
\subsection{Model Review}

We begin by briefly reviewing the unnatural composite Higgs model. Further details can be found in Ref.~\cite{Barnard:2014tla}. The underlying strong dynamics responsible for producing a composite Higgs has an $SU(7)$ global symmetry group which is spontaneously broken to $SU(6)\times U(1)$ at a scale $f\gtrsim 10$ TeV\@. This scale of breaking is chosen to satisfy all precision electroweak and flavour constraints without requiring any further symmetry in the model. This contrasts with the usual composite Higgs models where $f\gtrsim750$ GeV in order to minimise the tuning in the Higgs potential as much as possible, but where extra symmetries are needed to satisfy flavour and precision electroweak constraints.

The coset space $SU(7)/SU(6)\times U(1)$ contains twelve Nambu-Goldstone bosons which arrange themselves into a complex $\bf5$ of $SU(5)$ (containing the Higgs doublet, $H$, and a colour-triplet scalar, $T$) and a complex singlet, $S$. Note that this is the smallest coset space that preserves an $SU(5)$ symmetry and thus gauge coupling unification due to a composite top quark. The coset space also contains enough symmetry to prevent proton decay and stabilise the dark matter candidate, $S$. In particular, the strong sector is forbidden from mediating proton decay as it respects baryon number, $B$. It follows that it preserves baryon triality, a $\mathbb{Z}_3$ symmetry defined as
\begin{equation}
  \mathbb{Z}_B = 3B - n_C \mod 3 \,,
\end{equation}
where $n_C$ is the number of fundamental colour $(SU(3)_C)$ indices. All SM fields are neutral under this symmetry, while $T$ has $B(T) = B(H) = 0$ and $n_C = 1$.  Since a stable $T$ is trivially excluded, we must use baryon triality to stabilise $S$, achieved by arranging $B(S) = \frac{1}{3}$. A similar $\mathbb{Z}_3$ symmetry was previously used to stabilise composite fermionic dark matter in Refs.~\cite{Agashe:2004ci,Agashe:2004bm}.

The $SU(7)$ global symmetry is explicitly broken by coupling elementary-sector fields to composite-sector operators. This partial compositeness generates the Higgs potential whereupon a tuning, at least of order $10^{-4}$, is needed to obtain a 125 GeV Higgs boson. It also gives rise to masses for the singlet and colour-triplet scalars. The pseudo Nambu-Goldstone bosons $(H,T,S)$ are light ($\lesssim f$) and split from the composite-sector resonances which are much heavier ($\gg f$). There are also extra elementary-sector states, some of which are coloured, known as top companions. These are instrumental in decoupling the multiplet partners of the composite right-handed top quark and obtain a mass of order $f$. Thus the scalar triplet is the lightest, coloured exotic state predicted by the unnatural composite Higgs model and will generally be the most promising state to search for at colliders.

\subsection{Colour-Triplet Decay}

Because $T$ is charged under baryon triality ($\mathbb{Z}_B=+2$) it must decay to $S$, which has $\mathbb{Z}_B=+1$. Since the composite sector additionally preserves baryon and lepton number (required to forbid too-large neutrino masses) then the minimal possible decay is
\begin{equation}
  T \to u^c d^c S S\,,
\label{eq:Tdecay}\end{equation}
where $u^c$, $d^c$ are the $SU(2)$ singlet quarks with $\mathbb{Z}_B=0$. Further, it is clear that $t^c, b^c$ will dominate other final states, as the third generation couples most strongly to the composite sector. We would expect this decay to correspond to a dimension-6 operator in the low-energy effective theory after integrating out the heavy composite resonances. However, no such operator is generated in our model due to accidental symmetries associated with the necessary fermion representations. Instead, this decay is generated by the dimension-10 operator
\begin{equation}
  \lag \supset \frac{\Pi_3}{6\Lambda^4f^2} \, \lambda_{b^c} \lambda_\nu \lambda_\tau^\ast \;\epsilon_{i_3j_3k_3} \, (b^c)^{i_3} (t^c)^{j_3} (T^\dagger)^{k_3} \, S^2 \, l^\dagger /\!\!\!p\, l \,.
\label{eq:decayop}\end{equation}
Here, $\Pi_3 \sim 1 + \mathcal{O} (p^2/\Lambda^2)$ is a form factor, $\Lambda \approx g_\rho f$ is the approximate resonance mass, $g_\rho$ a strong sector coupling, and the $\lambda$'s are spurions for the partial compositeness couplings of the SM fermions. This operator exploits the fact that the lepton doublet has two couplings to the composite sector. It generates the decay of Eq.~\eqref{eq:Tdecay} after closing the leptons into a loop and this turns out to be less suppressed than the six-body final state.

Eq.~\eqref{eq:decayop} is only the leading contribution to the $T$ decay. Integrating out the composite sector will generate additional operators at higher orders. Further contributions to the decay~\eqref{eq:Tdecay} necessarily involve loops of elementary particles and are suppressed by $\lambda^2/(16\pi^2 g_\rho^2)$, where $\lambda$ is the appropriate elementary-composite spurion couplings. Other decay modes must involve at least two additional fermions, so are phase-space suppressed by $m_T^2/(8\pi\Lambda^2)$.  It is therefore a good approximation to neglect alternative operators.

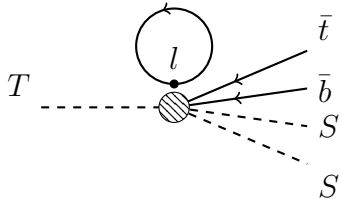
\begin{figure}
  \centering
  \begin{tikzpicture}[node distance=0.25cm and 1.75cm]
    \coordinate (v1);
    \coordinate[left = of v1, label=above left:$T$] (i1);
    \coordinate[vertex, above = of v1, label=above:$l$] (v2);
    \coordinate[above right = of v1, label= right:{$\bar{b}$}] (o2);
    \coordinate[above = of o2] (u1);
    \coordinate[above = of u1, label=above right:{$\bar{t}$}] (o1);
    \coordinate[below right = of v1, label= right:{$S$}] (o3);
    \coordinate[below = of o3] (u2);
    \coordinate[below = of u2, label=below right:{$S$}] (o4);
    \draw[scalar] (i1) -- (v1);
    \draw[fermion] (o2) -- (v1);
    \draw[fermion] (o1) -- (v1);
    \draw[scalar] (v1) -- (o3);
    \draw[scalar] (v1) -- (o4);
    \draw[fill = white] (v1) circle (0.2);
    \draw[fermion] (v2) arc (-90:270:0.5);
    \fill[pattern = north west lines] (v1) circle (0.2);
  \end{tikzpicture}
  \caption{Leading Feynman diagram for colour-triplet scalar decay.}\label{fig:MEdecay}
\end{figure}

The relevant Feynman diagram is shown in Figure~\ref{fig:MEdecay}. Neglecting the lepton mass the matrix element becomes
\begin{equation}
  i \ampM = - \frac{2i}{3\Lambda^4f^2} \, \lambda_{b^c} \lambda_\nu \lambda_\tau^\ast \;\epsilon_{i_3j_3k_3} \bar{u} (p_t) P_R u(p_b) \int \frac{d^4 p_l}{(2\pi)^4} \, (-1) \text{Tr} \biggl[ \frac{/\!\!\!p_l}{p_l^2} /\!\!\!p_l P_L \biggr] \Pi_3 \,,
\end{equation}
where $i_3,j_3,k_3$ are colour indices, $u,{\bar u}$ are spinors and $P_{L,R}$ are projection operators. The loop integral is cut off by the presence of composite resonances at the scale $\Lambda$.  We cannot compute this integral without knowledge of the physics at that scale, so we define
\begin{equation}
  \int \frac{d^4 p_l}{(2\pi)^4} \, (-1) \text{Tr} \biggl[ \frac{/\!\!\!p_l}{p_l^2} /\!\!\!p_l P_L \biggr]  \Pi_3 = - 2 \int \frac{d^4 p_l}{(2\pi)^4}  \Pi_3 = - 2 c^T_3 \frac{\Lambda^4}{(4\pi)^2}\,,
\end{equation}
where $c^T_3$ is an order-one constant. The matrix element now takes a simple form
\begin{equation}
  \frac{1}{3} \sum \lvert \ampM \rvert^2 = \biggl(\frac{c_3^T}{6\pi^2f^2} \biggr)^2 \lvert \lambda_{b^c} \lambda_\nu \lambda_\tau^\ast\rvert^2 \, 
  p_t \cdot p_b \,.\label{eq:squaredM}
\end{equation}
The calculation of the decay width is straightforward, though details regarding the four-body phase space integral are given in Appendix~\ref{app:phase space}. We define a dimensionless function, $J$, to capture the phase-space suppression from non-zero final state masses
\begin{equation}
  J(m_t, m_S) = \frac{72}{m_T^6} \int dQ_1^2 \, dQ_2^2 \, Q_1^2 \sqrt{I \biggl( \frac{Q_1^2}{m_T^2} ,  \frac{Q_2^2}{m_T^2} \biggr)} \, 
 \left( 1 - \frac{m_t^2}{Q_1^2} \right)^2  \sqrt{ 1 - \frac{4 m_S^2}{Q_2^2}} \,,
\end{equation}
where the function $I(a,b)$ is defined in Eq.~\eqref{eq:Idef}. The limits on the integrals are given by Eqs.~\eqref{eq:Q1bounds} and~\eqref{eq:Q2bounds}.  By construction, $J(0, 0) = 1$. The total width is
\begin{equation}
  \Gamma = \frac{(c_3^T)^2}{2^{19} 3^4 \pi^9} \, \lvert \lambda_{b^c} \lambda_\nu \lambda_\tau^\ast\rvert^2 \, \frac{m_T^5}{f^4} \, J(m_t, m_S) \,.
\end{equation}
Compared to the result in Ref.~\cite{Barnard:2014tla} the width in the zero-mass limit differs by a factor of 5/16.  Finally, making the replacements $\lambda_{b^c} \sim \sqrt{3 g_\rho y_b}$ and $\lambda_\nu \sim \lambda_\tau \sim \sqrt{2 g_\rho y_\tau}$, where $y_b (y_\tau)$ are the bottom (tau) Yukawa couplings, we obtain the approximate expression for the decay length
\begin{equation}
  c\tau = 0.6~\text{mm}\, \biggl( \frac{1}{c_3^T} \biggr)^2 \biggl( \frac{8}{g_\rho} \biggr)^3 \biggl( \frac{3~\text{TeV}}{m_T} \biggr)^5 \biggl( \frac{f}{10~\text{TeV}} \biggr)^4 \frac{1}{J(m_t, m_S)} \,.
\end{equation}
We see that for typical parameters in the unnatural composite Higgs model the decay length is of order the millimetre scale. The decay length can be substantially larger by either increasing the scale $f$, reducing the triplet mass, or having kinematic suppression $m_T \approx 2m_S + m_t$ (i.e. $J(m_t,m_S)\approx 0$). This behaviour is depicted in Figure~\ref{fig:decays} as a function of the colour-triplet and singlet scalar masses.

\subsection{Colour-Triplets in Other Unnatural Composite Higgs Models}

Any composite Higgs model that unifies via an $SU(5)$ gauge group will contain (at least) a colour-triplet pseudo Nambu-Goldstone boson like the one discussed here. Although other unification patterns are possible, precision unification in composite Higgs models is only known to occur via an $SU(5)$ gauge group, and only when the right-handed top quark is fully composite~\cite{Agashe:2005vg}. Unless a qualitatively different solution for precision unification is found light, colour-triplet scalars can therefore be considered a generic feature of unnatural composite Higgs models.

Whether the colour-triplet scalar is long-lived or not depends more on the details of the model. It will necessarily be charged under baryon triality, a symmetry that must hold at least approximately in order to prevent proton decay. This has a stabilising effect on the colour-triplet and means that it will preferentially decay via other exotic states. Furthermore, because the colour-triplet scalar is a pseudo Nambu-Goldstone boson the only available states are other pseudo Nambu-Goldstone bosons. In itself this is not enough to guarantee a long-lived state but, in the minimal model proposed in Ref.~\cite{Barnard:2014tla}, including the SM matter content resulted in several additional, accidental symmetries that stabilised the scalar colour-triplet even more. Accidental symmetries like these are increasingly likely to occur in more complicated models with larger initial symmetry groups so, while it is by no means certain, long-lived colour-triplet scalars seem likely to be a feature of most unnatural composite Higgs models exhibiting precision gauge coupling unification.

\section{Experimental Searches}\label{sec:ES}

We next discuss experimental searches for colour-triplet scalars. We first present limits from various direct searches that look for decays over a range of decay lengths. Afterwards we discuss indirect limits on the colour-triplet mass that arise from the modification of the Higgs couplings.

\subsection{R-hadron Searches}

ATLAS and CMS have published comprehensive R-hadron searches, including searches for charged R-hadrons escaping the detector~\cite{1305.0491,1411.6795} and searches for R-hadrons getting stopped by and then decaying within the detector~\cite{1310.6584,1501.05603}. The former analyses give rise to the strongest bounds so we shall use them to derive constraints on unnatural composite Higgs models, and also generalise them to estimate the R-hadron discovery and exclusion potentials of future experiments. Since our results depend only on the mass and width of the colour-triplet scalar they can be applied to any model predicting a long-lived particle of a similar nature.

The searches are characterised by low backgrounds, between zero and one event after 20~fb$^{-1}$ of 8~TeV collisions, and signal efficiencies around 10\%. The ATLAS study in Ref.~\cite{1310.6584} demonstrated that R-hadrons with more than 20~GeV of kinetic energy are not significantly slowed by the detector. We therefore take the following approach to derive constraints on unnatural composite Higgs models.
\begin{itemize}
\item Read in colour-triplet scalar production cross-sections from Ref.~\cite{1407.5066}.
\item Pair produce R-hadrons using the R-hadronisation routines in {\tt PYTHIA 8.1}~\cite{hep-ph/0603175,hep-ph/0611040,0710.3820}.
\item Discard R-hadrons with less than 20~GeV of kinetic energy.
\item Record the mass, energy, and transverse momenta of all remaining R-hadrons.
\item Weight each event by a survival factor (the probability of both R-hadrons escaping the detector).
\item Apply the reported signal acceptance-times-efficiency values.
\end{itemize}
In several of these steps we exploit the fact that the colour-triplet has the same quantum numbers as a (right-handed) sbottom, so various tools designed for SUSY searches can be easily repurposed.

Because the backgrounds are so low it is necessary to weight each event by a survival factor instead of allowing R-hadrons to decay directly in {\tt PYTHIA}\@. Prohibitively large numbers of events are otherwise needed to investigate the discovery and exclusion potentials of future experiments. The survival factor, $p$, for each R-hadron is given by
\be
p(r_{\rm decay}>r_{\rm detector})=e^{-\beta_Tr_{\rm detector}\Gamma/\gamma}
\ee
where $\beta_T$ is the R-hadron's transverse speed and $\gamma$ its overall Lorentz factor, both derived from the mass, energy, and transverse momentum of the R-hadron. $\Gamma$ is the colour-triplet width and we assume a value of $r_{\rm detector}=10$~m for the detector radius throughout this study.

For the number of background events we assume that the existing values will simply scale up with luminosity at future experiments. Taking a value from the ATLAS study in Ref.~\cite{1305.0491} gives 0.27~events per 19.1 fb$^{-1}$. Similarly, we assume that the signal acceptance-times-efficiency will remain constant, the same study giving a value of 0.084.

\begin{figure}[!h]
\begin{center}
\includegraphics[width=0.49\textwidth]{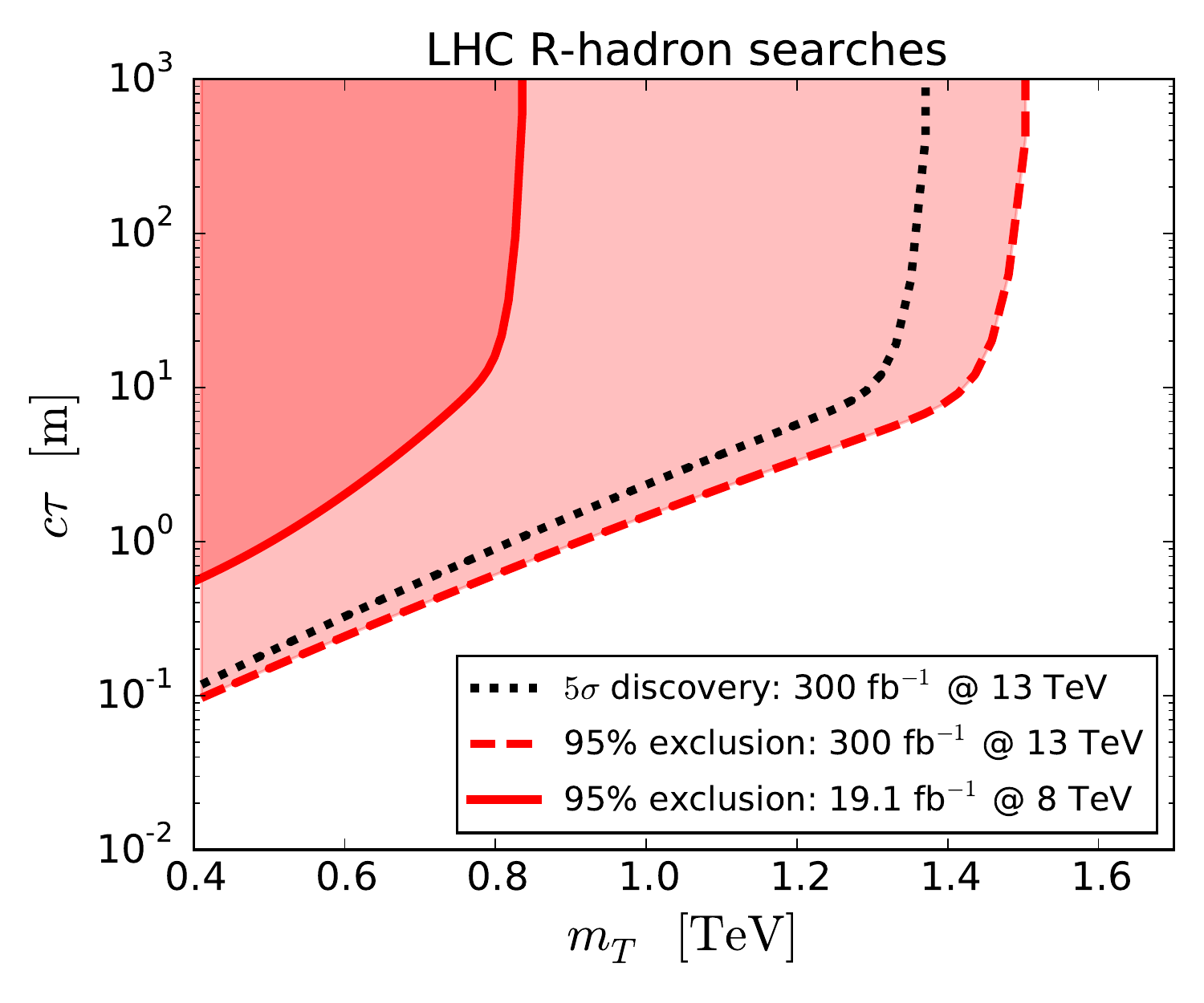}
\includegraphics[width=0.49\textwidth]{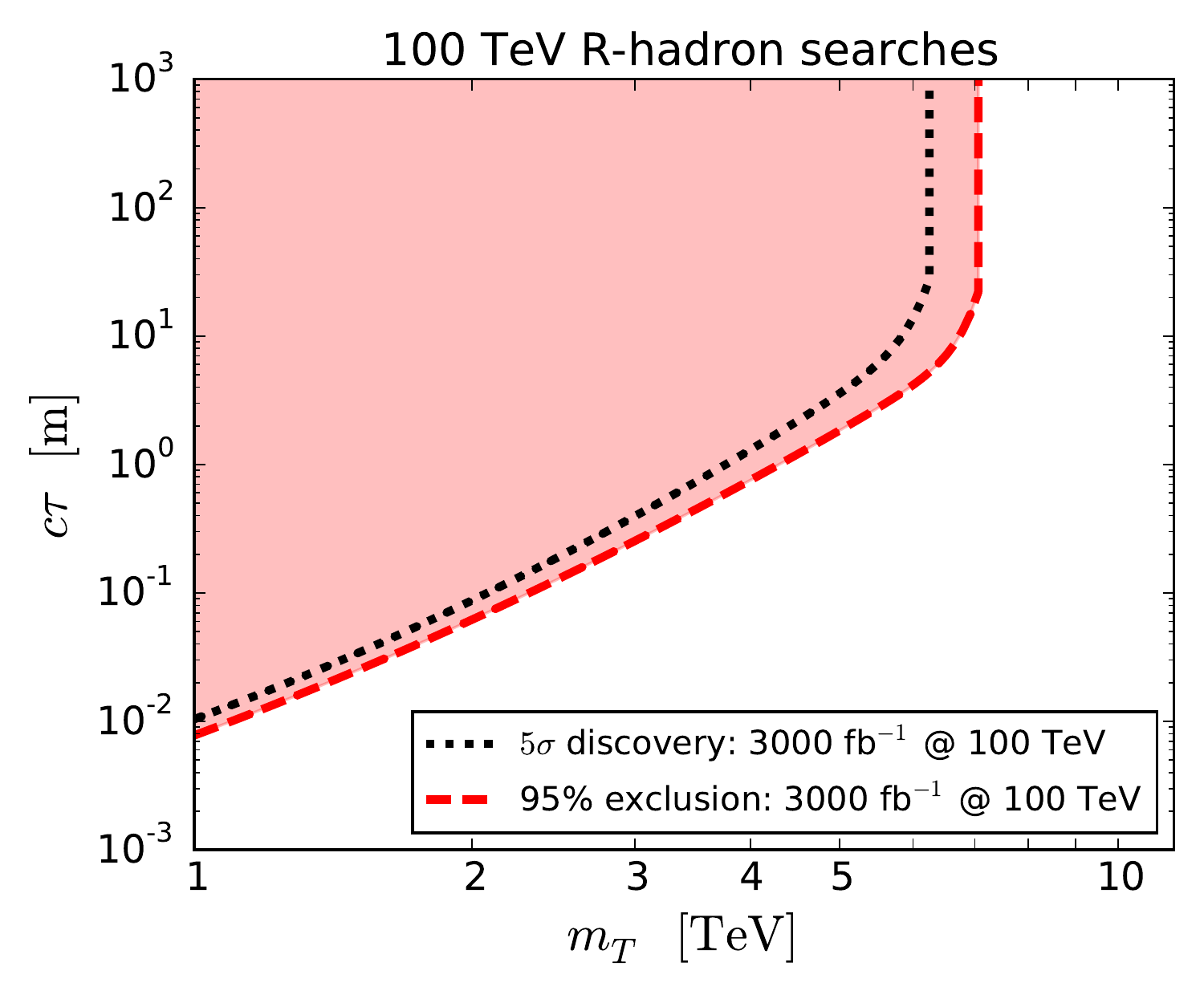}
\end{center}
\caption{Current status and future prospects for R-hadron searches as functions of colour-triplet scalar mass and lifetime.\label{fig:Rhad}}
\end{figure}

The results of this analysis are the discovery and exclusion contours shown in Figure~\ref{fig:Rhad}. These are presented in the plane of the colour-triplet mass, $m_T$, versus its lifetime, $c\tau$. We find that the final LHC dataset will be able to discover long-lived, colour-triplets with a mass up to around 1.4~TeV, and exclude those with a mass up to around 1.5~TeV\@. A 100~TeV collider would increase these values considerably, to 6 and 7~TeV respectively.

\subsection{Displaced-Vertex Searches}

Traditional heavy stable charged particle or R-hadron searches provide good sensitivity when the colour-triplet scalar is stable or has a long enough lifetime such that most of the decays occur outside the detector. However for shorter lifetimes these types of searches begin to lose sensitivity\footnote{ATLAS has now also performed a search for metastable R-hadrons~\cite{Aad:2015qfa} which decay within the detector at radial distances greater than 45~cm. This search is expected to have lower sensitivity than the displaced-vertex search we consider here except for a narrow range of lifetimes approaching the collider stable case.} as shown in Figure~\ref{fig:Rhad}. Dedicated searches for displaced decays are therefore essential in order to cover the entire parameter space of the model. There are now a variety of ATLAS~\cite{1310.3675,1409.0746,1501.04020,1504.03634,1504.05162} and CMS~\cite{1409.4789,1411.6530,1411.6977,CMS:2015gga} searches specifically targeting displaced signals. However recasting limits from these searches is difficult without access to the complete detector simulations used by the collaborations. Nevertheless several recent 
papers~\cite{1409.6729,1503.05923,1504.07293,1505.00784,1506.08206} have demonstrated that, with some reasonable assumptions, good agreement with the full experimental analyses can be achieved. The most relevant search for our model is the ATLAS displaced-vertex search~\cite{1504.05162} and we shall take a similar approach to that of Ref.~\cite{1505.00784}, which also reinterpreted this search but in the context of supersymmetric models with R-parity violation.

The ATLAS displaced-vertex search targets long-lived particles which decay within the inner detector, up to radial distances $\sim30$~cm. The search looks for displaced vertices containing at least five charged particle tracks in addition to the presence of a high-$p_T$ muon or electron, jets or missing energy ($\slashed{E}_T$). All channels are essentially background free with less than one event expected. We will focus on the DV+jets and DV+$\slashed{E}_T$ channels as these are expected to give the highest sensitivity to our colour-triplet decay. The displaced vertex requirements along with the final selection criteria in each of the channels are detailed in Table~\ref{tab:DV_selection}.

\begin{table}[h!]
  \centering
  \begin{tabular}{|c|c|}
    \hline
     & Selection criteria \\
    \hline
    \shortstack{displaced \\ vertex \\ \vspace{6mm}} & \shortstack{$\,$ \\ $\geq5$ tracks satisfying $p_T>1$~GeV, $|d_0|>2$~mm \\ DV position: $r_{DV}<300$~mm, $|z_{DV}|<300$~mm \\ and $\geq4$~mm from PV in transverse direction \\ $m_{DV}>10$~GeV (assuming $m_\pi^\pm$ for individual tracks)\\ material veto } \\   
    \hhline{|=|=|}
    \shortstack{DV+jets \\ $\,$} & \shortstack{$\,$ \\ $\geq4$ jets ($p_T>90$~GeV) or $\geq5$ jets ($p_T>65$~GeV) \\ or $\geq6$ jets ($p_T>55$~GeV) and $|\eta|<2.8$} \\
    \hline
    DV+$\slashed{E}_T$ & $\slashed{E}_T>180$~GeV \\
    \hline    
  \end{tabular}
  \caption{Displaced vertex requirements and final selection criteria for the ATLAS displaced-vertex search in the DV+jets and DV+$\slashed{E}_T$ channels.} \label{tab:DV_selection}
\end{table}

In replicating the experimental analysis we must also take into account the ATLAS tracking and vertex reconstruction procedures in addition to the above selections. The standard ATLAS tracking algorithms have a low efficiency for reconstructing tracks with large impact parameters ($d_0$, $z_0$) arising from displaced vertices. Therefore additional offline retracking is performed with looser requirements on $d_0$ and $z_0$. In order to account for this we have included an additional $|d_0|$-dependent efficiency factor multiplying the standard prompt efficiencies in the {\tt DELPHES 3}~\cite{deFavereau:2013fsa} detector simulation.

In simulating the ATLAS vertex reconstruction algorithm we adopt the same procedure as Ref.~\cite{1505.00784}. Firstly we consider only tracks with $p_T>1$~GeV, $|d_0|>2$~mm and truth-level origins satisfying $4<r<300$~mm and $|z|<300$~mm. Vertices are then reconstructed by firstly combining all track pairs with origins separated by $<1$~mm into a DV\@. The momentum vectors, $\vec{p}$, of these tracks must also satisfy $\vec{d}\cdot\vec{p}/|\vec{p}|>-20$~mm, where the vertex position, $\vec{d}$, with respect to the primary vertex (PV) is taken as the average position of its constituent track origins. Any vertices separated by $<1$~mm are then iteratively combined. Lastly vertices formed at radial distances corresponding to dense regions of the detector according to Ref.~\cite{1504.05162} are removed.

Finally we must also make some additional assumptions about how the long-lived R-hadrons and their decay products interact with and are reconstructed by the detector. This is particularly important in the case of the DV+$\slashed{E}_T$ channel in order to accurately estimate the missing energy. Firstly, we neglect any prompt tracks from R-hadrons that decay within the detector and which are anyway ignored when reconstructing displaced vertices. We also neglect the curvature of these R-hadron trajectories in the magnetic field, which will generally be small due to their large momenta. The decay products (excluding neutrinos) of R-hadrons decaying within the calorimeters are assumed to deposit all of their energy, although clearly this assumption is not expected to be valid for R-hadrons decaying near the outer edge. We neglect any energy deposits from the R-hadrons themselves which are expected to be small. R-hadrons decaying within the muon spectrometers are unlikely to be reconstructed as muons and are therefore assumed to contribute to $\slashed{E}_T$. Finally, charged R-hadrons which escape the detector are assumed to be reconstructed as muons.

\begin{figure}[h]
  \centering
  \includegraphics[width=0.6\textwidth]{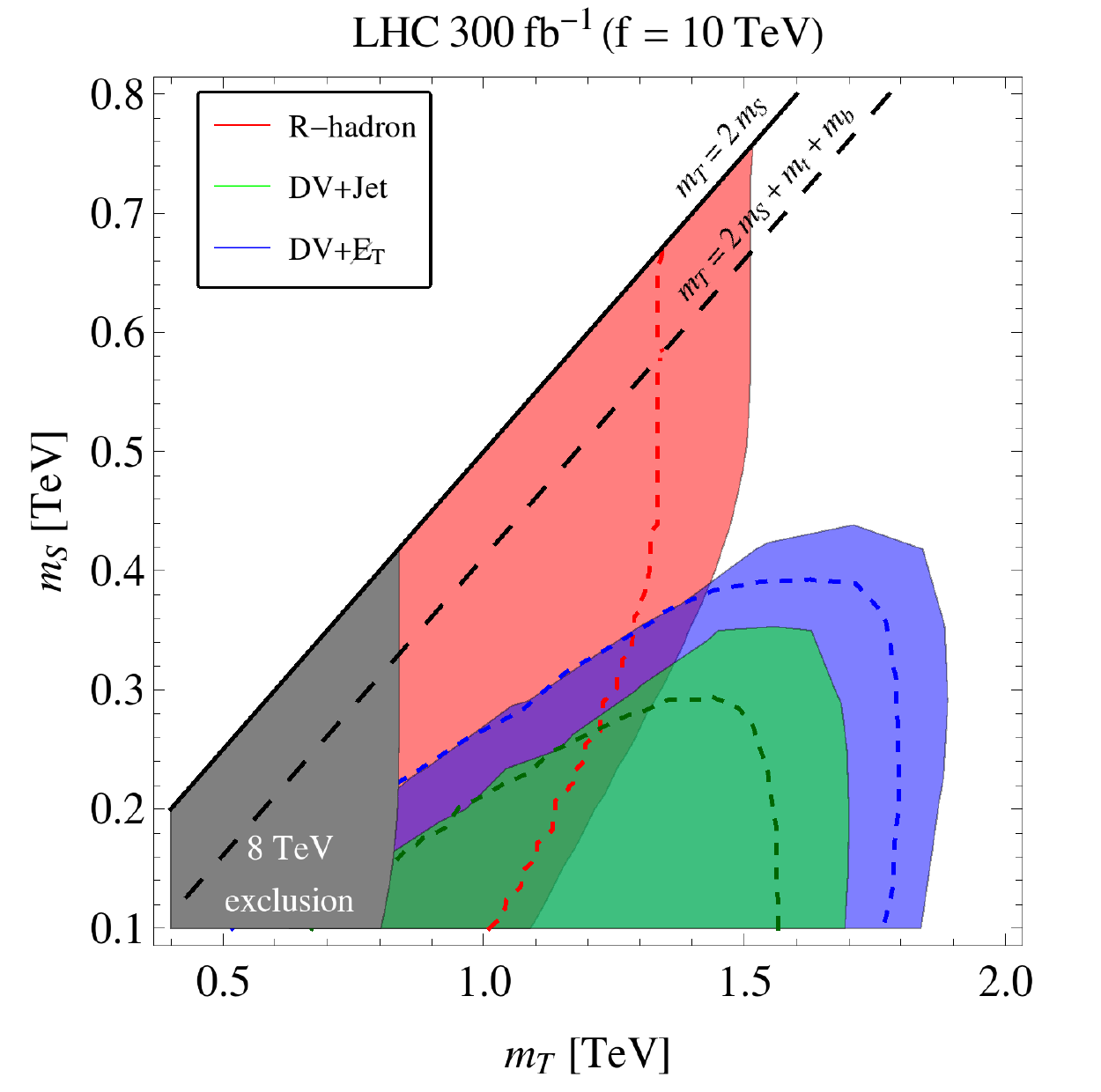}
  \caption{Projections for the R-hadron and displaced-vertex searches at the LHC with 300 fb$^{-1}$ of integrated luminosity at $\sqrt{s}=13$~TeV\@ as functions of the scalar mass $m_S$ and triplet mass $m_T$. The shaded regions can potentially be excluded at 95\% CLs and the dashed lines denote the $5\sigma$ discovery reach. The grey shaded region is excluded by current R-hadron searches at $\sqrt{s}=8$~TeV.}
  \label{fig:13TeVf10}
\end{figure}

\begin{figure}[h]
  \centering
  \begin{minipage}[b]{0.49\textwidth}
    \centering
  \includegraphics[width=\textwidth]{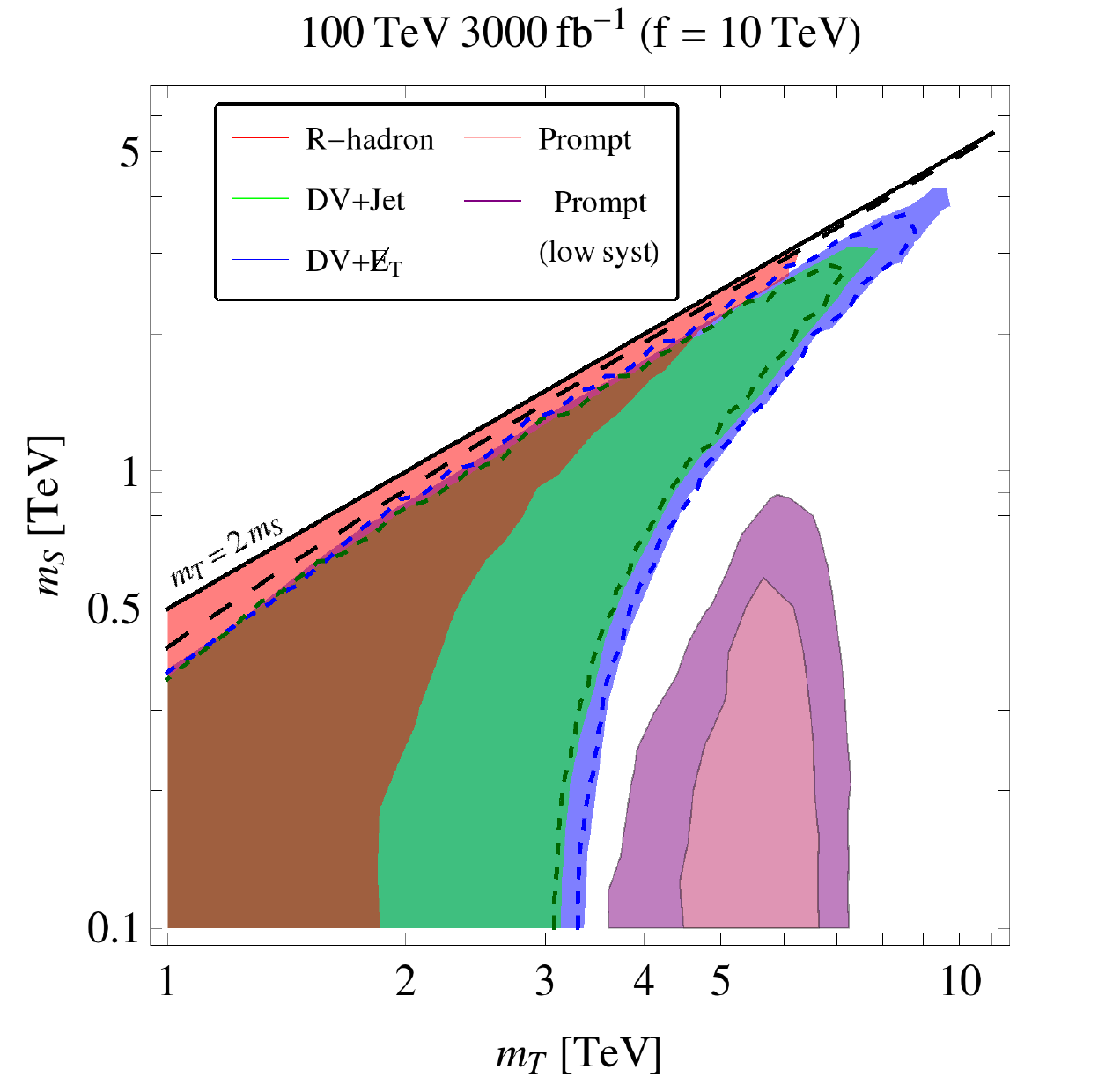}
  \end{minipage}
  \begin{minipage}[b]{0.49\textwidth}
    \centering
  \includegraphics[width=\textwidth]{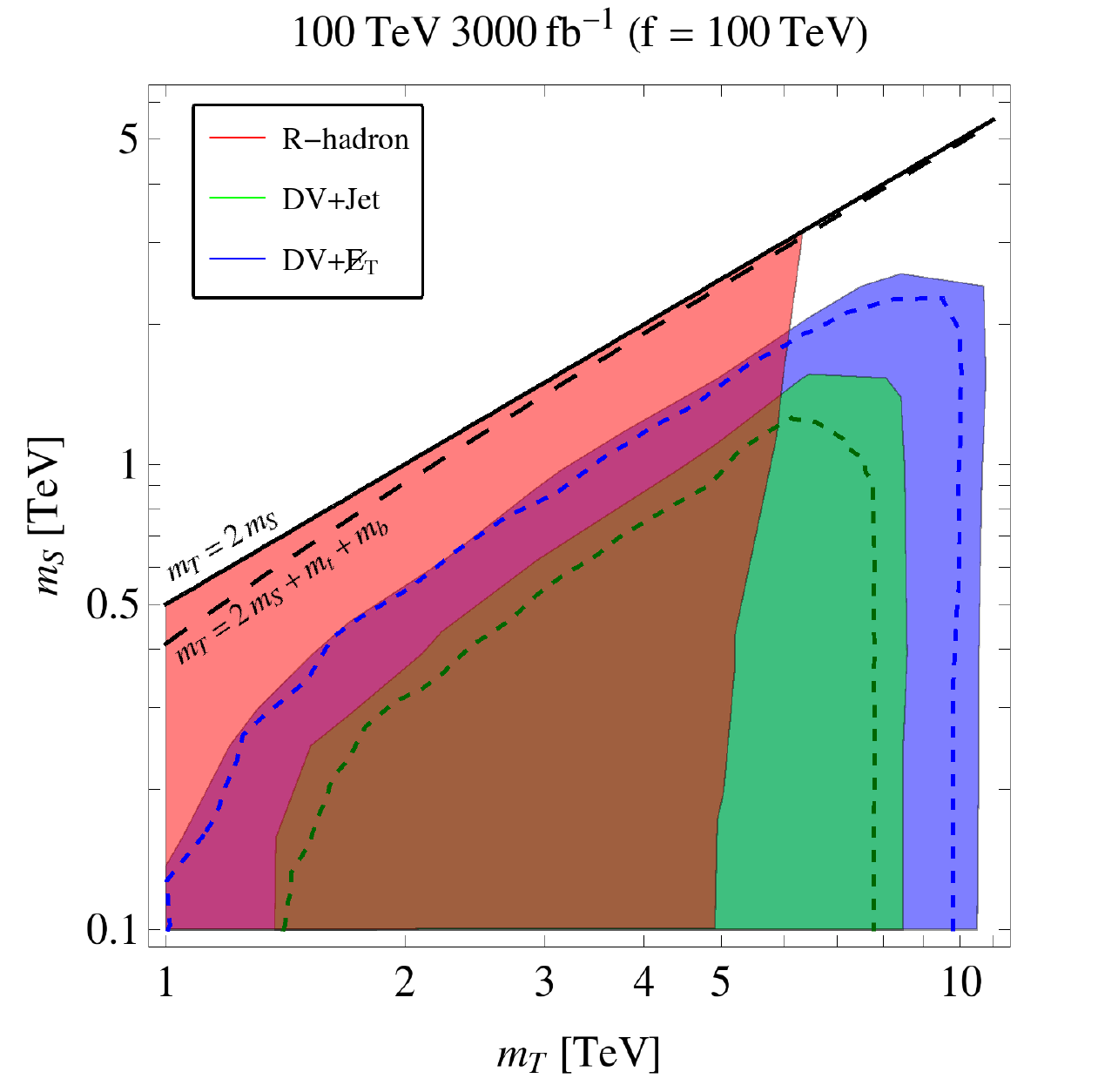}
  \end{minipage}
  \caption{Projections for a hypothetical $\sqrt{s}=100$~TeV collider with 3000 fb$^{-1}$ of integrated luminosity as functions of the scalar mass $m_S$ and triplet mass $m_T$. The shaded regions show the $5\sigma$ discovery reach (95\% CLs exclusion limit) for the R-hadron/displaced (prompt) searches. The dashed lines include an additional factor of two reduction in the signal efficiency for DV searches to account for the impact of more stringent experimental cuts. The left and right panels correspond to $f=10$ and 100~TeV respectively.} 
  \label{fig:100TeV}
\end{figure}

Similarly to the R-hadron search, signal events were generated using the R-hadronisation routines in {\tt PYTHIA} although with additional matrix-element re-weighting to correctly capture the kinematics of the 4-body decays of the triplet. The dominant (albeit very small) source of background for this search is due to low-$m_{DV}$ vertices which are crossed by an unrelated high-$p_T$ track. We assume that the current background expectations scale with increased luminosity while the systematic uncertainties remain fixed. We also assume a systematic uncertainty of 20\% on the signal efficiency. The $5\sigma$ discovery reach and 95\%~CLs exclusion limits in the ($m_T$, $m_S$) plane are then shown in Figures~\ref{fig:13TeVf10} and \ref{fig:100TeV}. Limits were computed in the {\tt ROOSTATS}~\cite{Moneta:2010pm} framework using the asymptotic formula for the profile likelihood~\cite{1007.1727} and Gaussian constraints for the systematic uncertainties.

We find that with the existing 8~TeV dataset this analysis does not have sufficient sensitivity to provide constraints on our colour-triplet scalar. This is due to the fact that for masses where the cross-section is sufficiently large the triplet is in most cases decaying outside the detector and R-hadron searches provide the only constraints. However displaced searches will become important to probe the full parameter space in Run-II and beyond. In Figure~\ref{fig:13TeVf10} we see that with 300~fb$^{-1}$ of integrated luminosity this search can potentially discover our colour-triplet up to masses of 1.8~TeV and exclude it up to 1.9~TeV\@. Furthermore this search is clearly complementary to the R-hadron searches considered in the previous section and the combination of both searches provides good coverage of the $(m_T, m_S)$ plane. For both searches the upper bound on the colour-triplet mass is cross-section limited and the reach is expected to improve with the increased dataset of the HL-LHC\@. Finally, we can also consider 
larger values of $f\gtrsim 100$ TeV, which increases the lifetime of the colour-triplet. In this case R-hadron searches will provide the only constraints at the LHC\@. 

In Figure~\ref{fig:100TeV} we also consider the prospects for this search at a hypothetical $\sqrt{s}=100$~TeV collider. We have assumed the same experimental cuts as the current ATLAS analysis, which leads to signal efficiencies of up to $\sim70\%$ for the highest colour-triplet masses considered. Of course in practice the cuts are likely to be more stringent, driven either by trigger considerations or background expectations derived from data. Although note that the signal efficiency can reach $60\%$ for some of the benchmark models considered in the existing analysis, suggesting that our estimate is not unreasonable. Nevertheless we also show results with the signal efficiency reduced by a factor of two in order to provide a more conservative estimate of the discovery reach. Regardless, we find that the reach would be significantly greater than at the LHC with potential discovery of the scalar triplet up to masses around 10~TeV\@.

\subsection{Prompt Decay Searches}

Standard searches for prompt decays of the colour-triplet are not expected to provide useful constraints at the LHC\@. This is simply due to the fact that for masses below about 4~TeV (assuming\footnote{For larger values, $f\gtrsim100$~TeV, prompt-decay searches will not be constraining even at a future $\sqrt{s}=100$~TeV collider and all limits will be from displaced-vertex and R-hadron searches.} $f=10$~TeV) most of the colour-triplet decays will be displaced, while for higher masses the LHC will not produce enough events even by the end of the planned HL-LHC upgrade. However future colliders may be able to probe this region of parameter space where the colour-triplet lifetime is small enough to lead to prompt decays, less than about 2~mm. 

We therefore investigate the potential limits from a hypothetical 100~TeV proton collider. Of course many assumptions have to be made about the future performance of such a machine and we will use the Snowmass detector~\cite{Anderson:2013kxz} implemented in {\tt DELPHES} to model the detector performance. We also make use of the Snowmass background Monte-Carlo event samples~\cite{Avetisyan:2013onh}. Signal events were again generated using {\tt PYTHIA} and we use the same weighted event generation procedure as used for the background events in order to obtain a sample suitable for studies with high integrated-luminosity. In our case the events are separated in bins of $p_T$ to allow for straightforward implementation using {\tt PYTHIA} and $50\,000$ events are generated in each bin.

The ATLAS experiment has recently performed a search for gluinos~\cite{Aad:2014lra} which considers a similar final state to that which arises from the pair production of our colour-triplet. We will employ a similar search strategy for our 100~TeV analysis, however extracting the signal for the colour-triplet case is significantly more challenging due to the reduced cross-section and, as we shall see, this leads to a relatively limited reach even at $\sqrt{s}=100$~TeV\@. We will focus on a search using the purely hadronic final state. Searches in the leptonic channel were also considered but are expected to be less sensitive for higher triplet masses due to the small cross-sections combined with a lower branching fraction. To begin we make the following preselection cuts:

\begin{itemize}
  \item $\geq4$~jets with $p_T>50$ GeV, $|\eta|<2.5$,
  \item $\geq3$~b-tagged,
  \item leading jet $p_T>150$~GeV,
  \item $\delta\phi^{4j}_{min}>0.5$,
  \item $\slashed{E}_T>400$ GeV,
  \item $m_\mathrm{eff}>2000$~GeV,
  \item No isolated leptons ($p_T>20$~GeV, $|\eta|<2.5$).
\end{itemize}
Jets are reconstructed using the anti-$k_T$ algorithm~\cite{Cacciari:2008gp,Cacciari:2011ma} with $R=0.5$ and we use the Snowmass loose b-tagging working point with a b-tag efficiency of 70-75\% and a light quark (c-quark) mis-tag rate of 3\% (30\%). $\delta\phi^{4j}_{min}$ is defined as the minimum azimuthal separation between $\slashed{E}_T$ and each of the four leading jets with $p_T>20$ GeV and $|\eta|<4.5$. The cut on this variable is designed to reduce the contribution to $\slashed{E}_T$ from poorly reconstructed jets or neutrinos emitted in the direction of a jet. Combined with the cut on $\slashed{E}_T$ this is expected to reduce the QCD background to a negligible amount, although the QCD background has not been simulated as part of the background sample. Finally, $m_\mathrm{eff}$ is defined as the scalar sum of $\slashed{E}_T$ and all jets with $p_T>50$ GeV and $|\eta|<4.5$. We also neglect events where the triplet decay vertex is displaced by more than 2~mm in the radial direction since they would likely fail b-tagging track requirements~\cite{ATL-PHYS-PUB-2015-022,Jindal:2012wra}.

\begin{figure}[h]
  \centering
  \begin{minipage}[b]{\textwidth}
    \centering
    \includegraphics[width=0.6\textwidth]{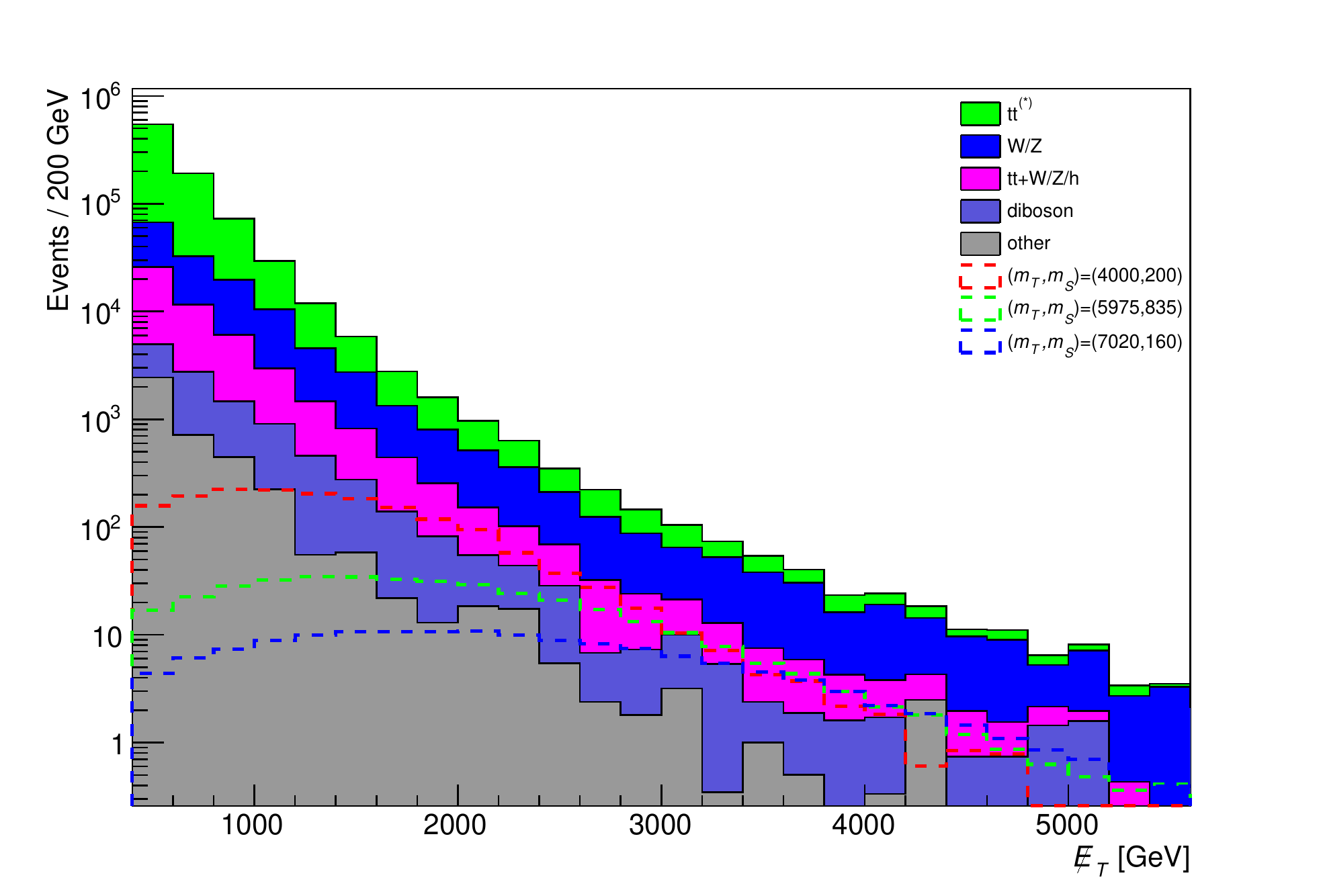}
  \end{minipage}
  \begin{minipage}[b]{\textwidth}
    \centering
    \includegraphics[width=0.6\textwidth]{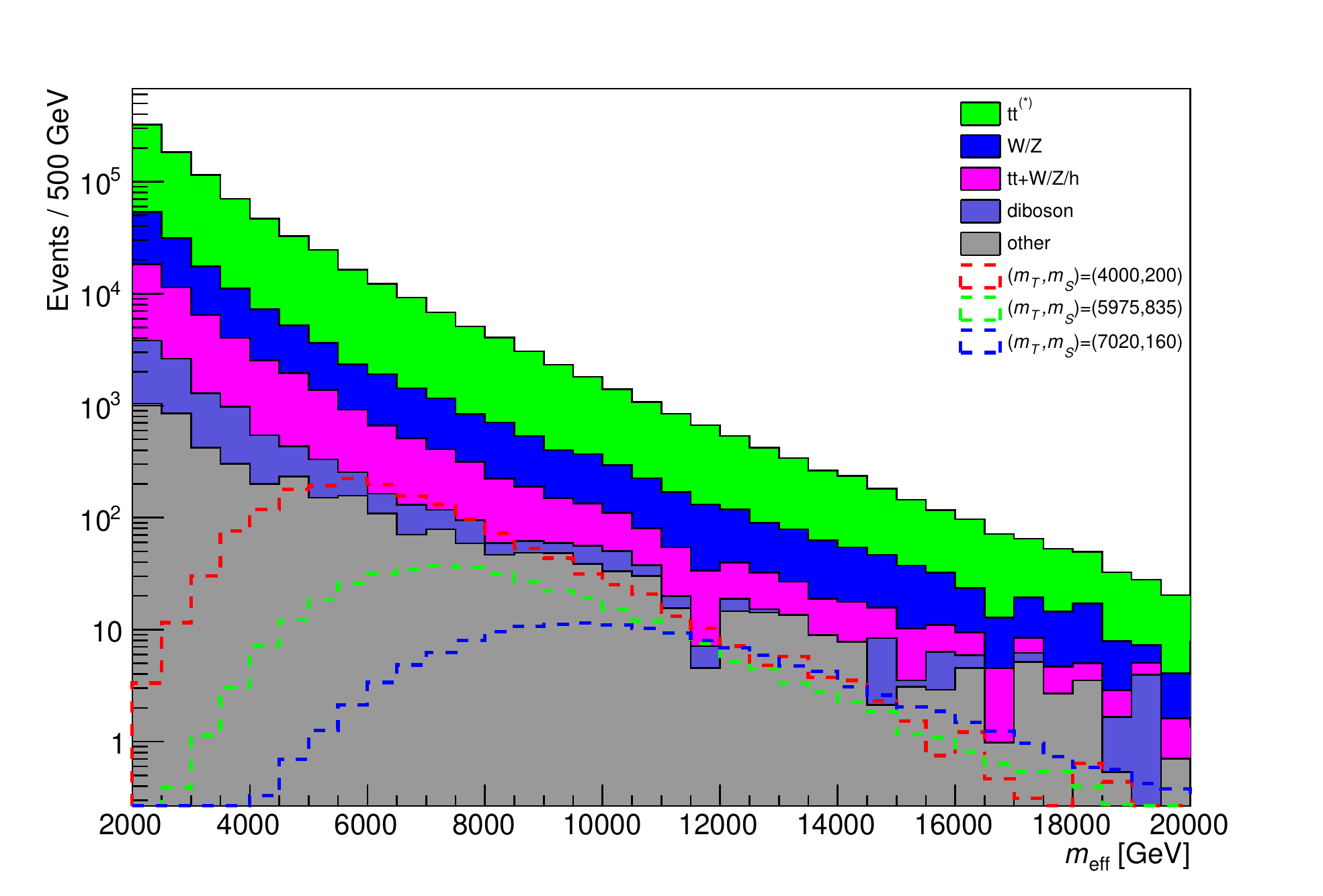}
  \end{minipage}
  \caption{The $\slashed{E}_T$ (upper) and $m_\mathrm{eff}$ (lower) distributions for the backgrounds and three benchmark signal points after imposing the preselection cuts.} \label{fig:prompt_distributions}
\end{figure}

After these preselection cuts the background still dominates over the signal in the selected sample by several orders of magnitude. The dominant background for this search is $t\bar{t}\,+$~jets. While we expect our signal to exhibit a higher b-jet multiplicity and increased $\slashed{E}_T$ compared to the background, the large $t\bar{t}$ cross-section means that the number of background events can still easily exceed the signal expectation even in the tails of the background distributions. This can be clearly seen in Figure~\ref{fig:prompt_distributions} where we have plotted the signal and background distributions of $\slashed{E}_T$ and $m_\mathrm{eff}$ after applying the preselection cuts for three benchmark signal points.

Next we optimise the cuts\footnote{Additional cuts on the number of jets and leading jet $p_T$ were also considered but found not to provide significant improvement in the background rejection.} on the number of b-jets ($N_b$), $\slashed{E}_T$ and $m_\mathrm{eff}$ in order to obtain the optimal background rejection as a function of signal efficiency using the {\tt TMVA} package~\cite{Hocker:2007ht} in {\tt ROOT v5.34}. This was performed separately for each signal point in a scan over the $(m_T, m_S)$ plane. However we find that the cuts yielding the maximum signal significance do not vary significantly over the parameter ranges of interest. We therefore impose the following final cuts when deriving the exclusion limits: $N_b\geq4$, $\slashed{E}_T>2.5$~TeV, $m_\mathrm{eff}>10$~TeV\@. The background and signal yields for three benchmark points after imposing the preselection and final cuts are shown in Table~\ref{tab:prompt_selection}.

\begin{table}[h!]
  \centering
  \begin{tabular}{|c|c|c|}
    \hline
    & Preselection  & \shortstack{$\,$ \\ Final selection \\ ($N_b\geq4$, $\slashed{E}_T>2.5$~TeV, $m_\mathrm{eff}>10$~TeV)} \\
    \hline
    $t\bar{t}^{(*)}\,+$~jets & $7.2\times10^5$ & 27 \\
    \hline
    $W/Z\,+$~jets & $9.1\times10^4$ & 10 \\
    \hline
    $t\bar{t}\,+\,W/Z$ & $3.9\times10^4$ & 3.8 \\
    \hline
    Other & $1.1\times10^4$ & 1.7 \\
    \hline    
    \hhline{|=|=|=|}
    Total background & $8.6\times10^5$ & 39 \\
    \hhline{|=|=|=|}
    \shortstack{$\,$ \\ $m_T=4000$~GeV \\ $m_S=200$~GeV} & 1720 & 13 \\
    \hline
    \shortstack{$\,$ \\ $m_T=5975$~GeV \\ $m_S=835$~GeV} & 378 & 19 \\
    \hline
    \shortstack{$\,$ \\ $m_T=7020$~GeV \\ $m_S=160$~GeV} & 147 & 22 \\
    \hline    
  \end{tabular}
  \caption{Background and signal event yields before and after the final selection for three benchmark signal points.} \label{tab:prompt_selection}
\end{table}

We can now compute 95\% CLs exclusion curves in the $(m_T, m_S)$ plane. The following systematic uncertainties are assumed in computing the limits: background normalisation (20\%), signal efficiency (15\%), PDF (5\%) and luminosity (2.8\%). We also consider the more optimistic assumption of 10\% and 5\% systematics for the background normalisation and signal efficiency respectively\footnote{With reduced systematic uncertainties the analysis does benefit from additional signal regions (e.g.\ $\slashed{E}_T>1.8$~TeV, $m_\mathrm{eff}>6$~TeV) targeting the low mass region. In this case we derive our exclusion limits using the optimal cuts for each $(m_T, m_S)$ point.}. The final exclusion curves are shown in Figure~\ref{fig:100TeV}. We see that for the lowest singlet masses we are able to potentially exclude triplet masses in the range 4-7~TeV\@. This upper reach is consistent with previous studies of colour-triplets at $\sqrt{s}=100$~TeV colliders in the context of supersymmetric simplified models~\cite{Cohen:2013xda}. However note that in our scenario there is no region in the $(m_T, m_S)$ parameter space where we are able to achieve a 5$\sigma$ discovery potential. One might expect this to be attainable for lower masses, where the cross-section is larger, however the colour-triplet then becomes long-lived and we must turn instead to displaced searches for the strongest limits. Once again this search is clearly complementary to the R-hadron and displaced-vertex searches and all three search strategies will be essential in order to probe the entire $(m_T, m_S)$ plane. Although we see from Figure~\ref{fig:100TeV} that there remains a narrow region between the prompt and displaced regimes which may be challenging to explore.

Finally, there are inevitably many assumptions which must be made in estimating the reach of future colliders. The analysis considered here relies heavily on b-tagging and this is likely to provide the largest source of uncertainty. We have chosen to use the loose b-tagging point defined for the Snowmass detector in our analysis as this assumes a reasonably conservative estimate on the mis-tag rate of 3\%. Improvements in b-tagging at the LHC have demonstrated that this kind of performance is reasonable for both highly boosted jets~\cite{ATL-PHYS-PUB-2014-014} and in high pile-up environments~\cite{Jindal:2012wra,ATL:upgrade}. We have also neglected the effects of pile-up in our analysis, however we do not expect this to have a significant effect beyond the impact on b-tagging. The assumptions made about the systematic uncertainties also have a significant effect on the final exclusion limit.

\subsection{Higgs Loop Decays}

Most corrections to the SM Higgs properties in composite Higgs models scale like $v^2/f^2$, and as such are unobservable given our lower bound $f\gtrsim 10$ TeV.  There are two possible exceptions to this rule: loop contributions from other Nambu-Goldstone bosons, and the Higgs coupling to $Z\gamma$.  Loop corrections from the scalar triplet will scale as $v^2/m_T^2$.  We have already considered $m_T \ll f$ in the previous sections; in such cases the triplet contributions could be substantially enhanced.  Limits derived this way are also independent of the triplet decay mode.  Second, the $hZ\gamma$ coupling is unique in being loop-level in the SM yet allowed by the shift symmetry~\cite{1308.2676} (through the operator $\gamma^{\mu\nu} Z_\mu \partial_\nu h$).  
This could potentially allow a contribution enhanced by strong sector couplings $g_\rho/g_{SM}$ and large numerical multiplicities.

\begin{figure}
  \centering
  \includegraphics[width=0.48\textwidth]{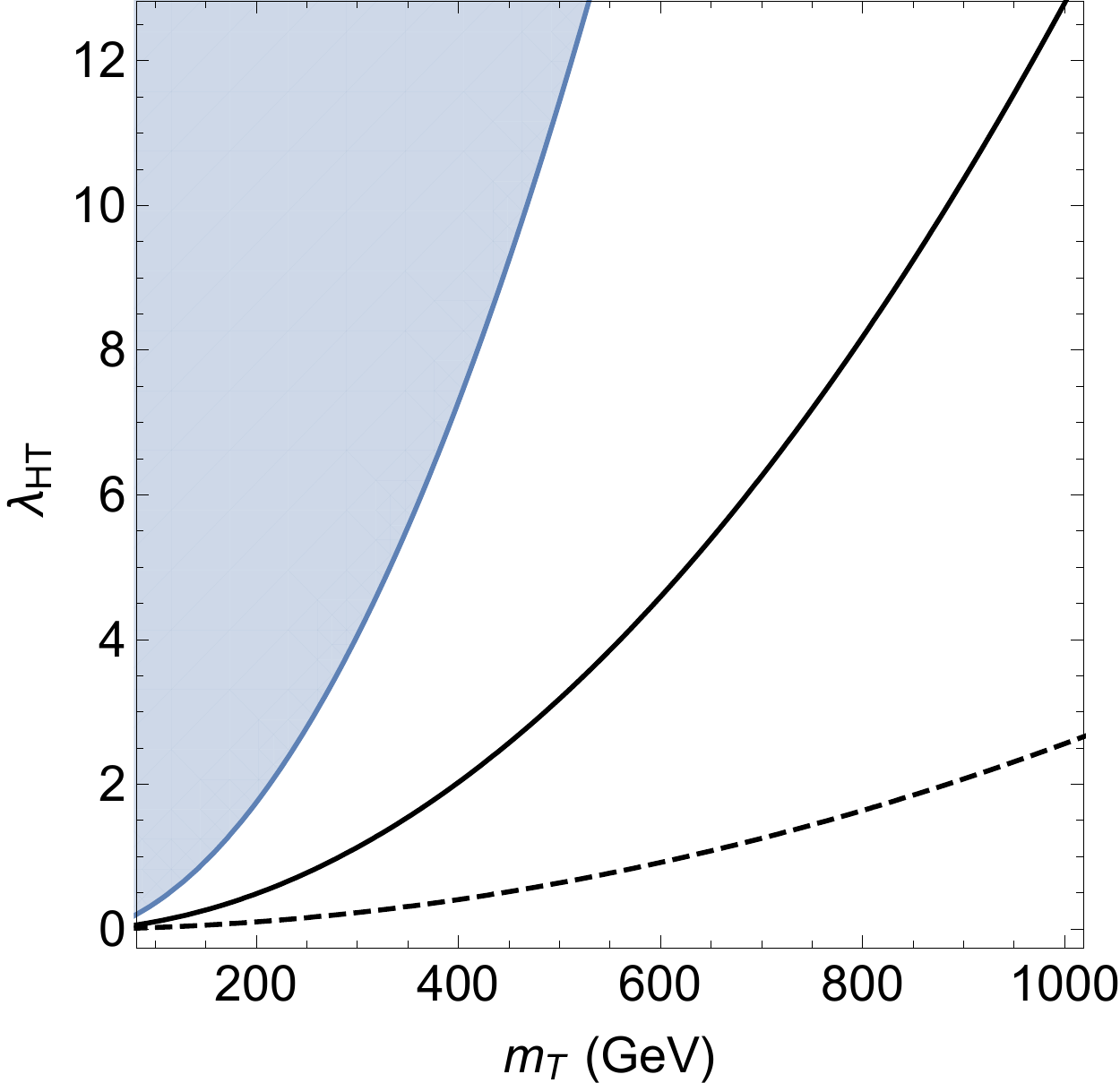}
  \includegraphics[width=0.48\textwidth]{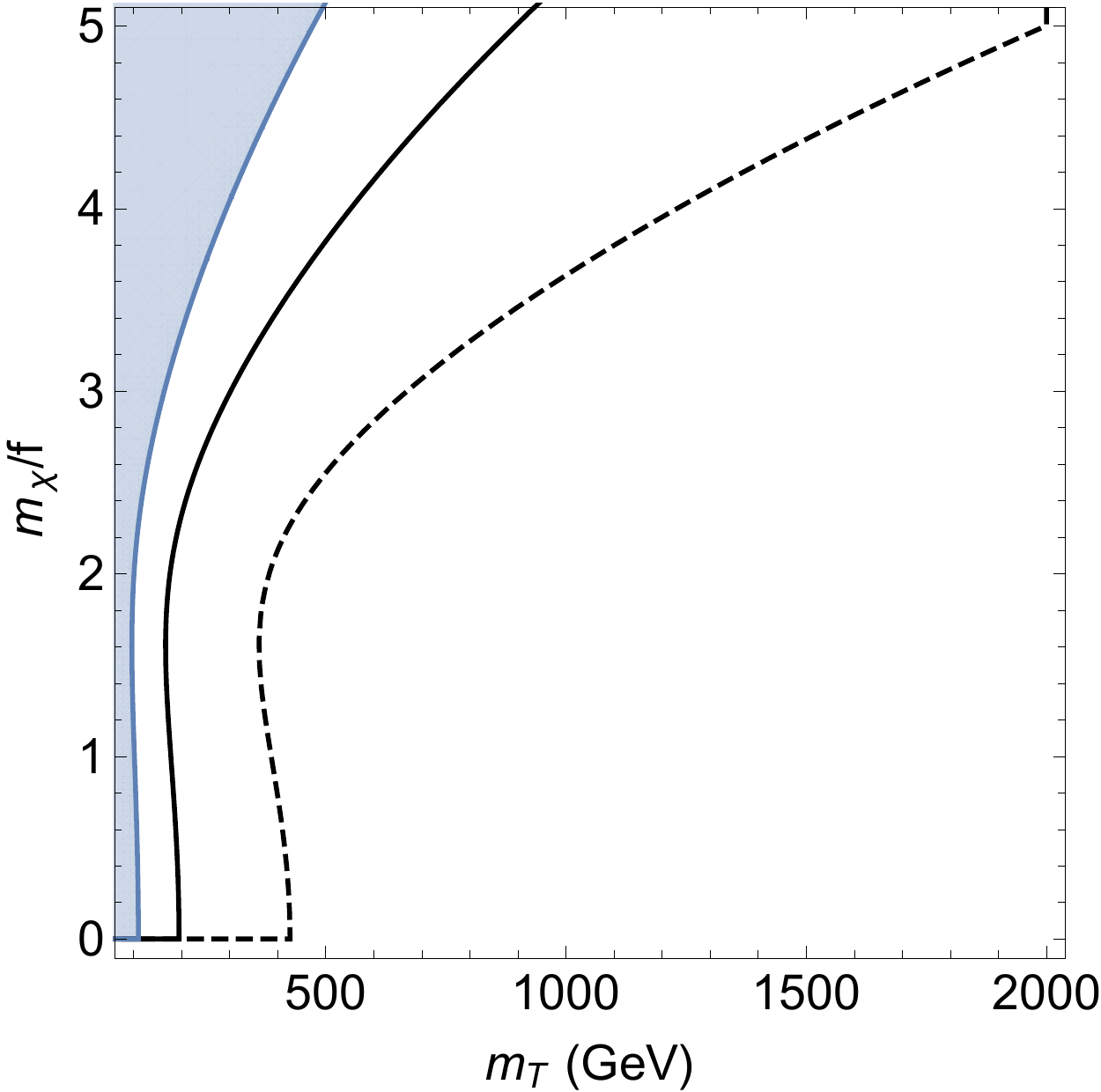}
  \caption{The triplet mass $m_T$ regions excluded by the Higgs coupling to gluons as functions of the Higgs-triplet quartic coupling, $\lambda_{HT}$ (left) and top companion coupling, $\lambda_\chi = m_\chi/f$ (right). The shaded regions are excluded by the current LHC measurements and the solid (dashed) lines show the prospective exclusions from the HL-LHC and the ILC.}\label{fig:excl}
\end{figure}

The modifications to the Higgs coupling from loops of new particles are well-known (see~\cite{Gunion:1989we} and references therein). We follow Ref.~\cite{1307.1347} in parameterising the shifts in the $h\gamma\gamma$, $hgg$, and $hZ\gamma$ couplings in terms of effective scale factors $\kappa_i$
\begin{equation}
  \kappa_{g,\gamma,Z\gamma} = 1 + \frac{\Delta \ampA_{g,\gamma,Z\gamma}}{\ampA_{g,\gamma,Z\gamma}^{SM}} \,,
\end{equation}
where $\Delta \ampA_{i}$ ($\ampA_{i}^{SM}$) is the new physics (SM) contribution to the loop. For the colour-triplet scalar, $T$, the contributions to the photon and gluon decays are very similar
\begin{align}
  \Delta\ampA_{g} & = \frac{\lambda_{HT} v^2}{2m_T^2} \, A_0 (\tau_h) \,, & \ampA_g^{SM} & \approx 1.3\,, \\
  \Delta\ampA_{\gamma} & = \frac{\lambda_{HT} v^2}{3m_T^2} \, A_0 (\tau_h) \,, & \ampA_\gamma^{SM} & \approx - 13\,, \\
  A_0 (\tau) & = - \tau \bigl[ 1 - \tau f(\tau) \bigr]\,, & f(\tau) & = \arcsin^2 \sqrt{1/\tau} \text{ (if } \tau > 1)\,,\\
  \tau_h & = \frac{4 m_T^2}{m_h^2}\,, & v & = 246~\text{GeV}\,,
\end{align}
and $\lambda_{HT}$ is the scalar quartic term in the potential
\begin{equation}
  \lag \supset \lambda_{HT} \, H^\dagger H \, T^\dagger T \,.
\label{eq:htquartic}\end{equation}
The contribution to the $Z\gamma$ decay is slightly more complex
\begin{align}
  \Delta\ampA_{Z\gamma} & = \frac{\lambda_{HT} v^2}{3 s_W c_W m_T^2} \, I_1 (\tau_h, \tau_Z) \,, & \ampA_{Z\gamma}^{SM} & \approx 1,
\end{align}
where $s_W$ ($c_W$) is the sine (cosine) of the Weinberg angle, and the loop function
\begin{align}
  I_1 (a, b) & = \frac{a b}{2(a - b)} + \frac{a^2 b^2}{2(a-b)^2} \bigl(f(a) - f(b)\bigr) + \frac{a^2 b}{(a-b)^2} \bigl( g(a) - g(b)\bigr) \,, \\
  g (\tau) & = \sqrt{\tau - 1} \, \arcsin \sqrt{1/\tau} \text{ (if } \tau > 1)\,, \qquad \tau_Z = \frac{4m_T^2}{m_Z^2} \,.
\end{align}
The current bounds from ATLAS, assuming no corrections to the other Higgs couplings, are~\cite{ATLAS:2015bea}
\begin{equation}
  \kappa_\gamma = 1.00 \pm 0.12\,, \qquad \kappa_g = 1.12 \pm 0.12 \,, \qquad  \kappa_{Z\gamma} < 3.3 \,.
\end{equation}
It is clear that $T$ shifts $\kappa_g$ much more than $\kappa_\gamma$, due to the relative size of the SM contributions. The shift to $\kappa_{Z\gamma}$ is also subdominant due to a cancellation in the loop function. We show the 95\% exclusion contour in the mass quartic-coupling plane in the left panel of Figure~\ref{fig:excl}. We also show projected limits from the HL-LHC and ILC from $\kappa_g$, assuming a SM central value and uncertainties of 5\%~\cite{1509.08721,ATL-PHYS-PUB-2014-016,1307.7135} and 1\%~\cite{1310.0763,1310.8361} respectively. These results hold for the generic case where a colour-triplet scalar is the only new light coloured state coupling to the Higgs.

In the unnatural composite Higgs the quartic coupling~\eqref{eq:htquartic} is calculable up to order-one coefficients~\cite{Barnard:2014tla}
\begin{align}
  \lambda_{HT} = \frac{1}{16\pi^2} \, \biggl( & \frac{28}{9}c{}_2^{\chi\chi}|\lambda_\chi|^4+\frac{8}{9}c{}_2^{tt}|\lambda_t|^4+\frac{4}{9}c{}_2^{bb}|\lambda_b|^4+\frac{8}{9}c{}_2^{b^cb^c}|\lambda_{b^c}|^4 \nonumber\\
& -\frac{16}{9}c{}_2^{\chi t}|\lambda_\chi|^2|\lambda_t|^2+\frac{2}{3}c{}_2^{tb}|\lambda_t|^2|\lambda_b|^2-\frac{4}{3}c{}_2^{bb^c}|\lambda_b|^2|\lambda_{b^c}|^2 \biggr) \,.
\end{align}
The consequent exclusions in terms of $m_\chi/f \approx \lambda_\chi$, where $m_{\chi}$ is the mass of the top companions, are shown in the right panel of Figure~\ref{fig:excl}, assuming $\lambda_t \approx 3 y_t$ (where $y_t$ is the top Yukawa coupling), $\lambda_{b,b^c} \approx \sqrt{3 g_\rho y_b}$, $g_\rho \approx 8$, and all $c_2^i \approx 1$. The contributions from the bottom Yukawa are negligible. There is a model-independent limit from the top Yukawa, which from the LHC is $m_T \gtrsim 100$~GeV\@. For heavy or light top companions the limits gets stronger, with the ILC able to exclude 2~TeV triplets for top companions with masses $m_\chi = 5 f$. In addition to the limits of the previous section these should be compared to the bound $m_T > 2m_S$ required to avoid a stable colour-triplet.  The current LUX bounds~\cite{Akerib:2013tjd} enforce $m_S \gtrsim 150$~GeV and hence $m_T \gtrsim 300$~GeV, which is already superior to potential HL-LHC bounds unless $m_\chi \gtrsim 3f$.

Finally, we note that in the unnatural composite Higgs model the contributions to the $hZ\gamma$ coupling from the strongly-interacting sector vanish at leading order.  This is a consequence of the unbroken $SU(5)$ global symmetry, and thus applies to any composite Higgs model compatible with an $SU(5)$ or $SO(10)$ GUT.  It is distinct from the parity argument discussed in Ref.~\cite{1308.2676}, as that symmetry only exists in models with a custodial $SU(2)$.  The low-energy effective theory for the Nambu-Goldstone bosons in the absence of explicit breaking of the global symmetry is given by the CCWZ expansion.  Because this respects $SU(5)$, the gauge fields can only appear in two forms: as part of the Nambu-Goldstone covariant derivatives, and in the $SU(5)$ matrix form
\begin{equation}
  F_{\mu\nu} = 
  \begin{pmatrix}
    g_s G_{\mu\nu}^a t_{SU(3)}^a - \frac{1}{3} g' B_{\mu\nu} \mathbf{1}_{3\times 3} & 0 \\
    0 & \frac{1}{2} (g W_{\mu\nu}^i \sigma^i + g' B_{\mu\nu} \mathbf{1}_{2\times 2})
  \end{pmatrix} \,,
\label{eq:su5gauge}\end{equation}
where $t_{SU(3)}^a$ are the Gell-Mann matrices and $\sigma^i$ are the Pauli matrices.  The lower block diagonal term is the one which multiplies the Higgs field when $F_{\mu\nu}$ is contracted with the Nambu-Goldstone field.  In terms of mass-basis fields, we have 
\begin{equation}
  F_{\mu\nu}^{(2)} \equiv \frac{1}{2} \left(g W_{\mu\nu}^i \sigma^i + g' B_{\mu\nu} \mathbf{1}_{2\times 2}\right) \sim 
  \begin{pmatrix}
    \gamma_{\mu\nu} & W_{\mu\nu} \\
    W^\dagger_{\mu\nu} & Z_{\mu\nu} 
  \end{pmatrix} \,.
\end{equation}
There are only three possible terms that can appear at dimension-6 involving the Higgs and gauge fields:
\begin{equation}
  H^\dagger F^{(2)}_{\mu\nu} F^{(2)\mu\nu} H \,, \qquad (D^\mu H)^\dagger F^{(2)}_{\mu\nu} D^\nu H \,, \qquad \epsilon^{\mu\nu\rho\sigma} H^\dagger F^{(2)}_{\mu\nu} F^{(2)}_{\rho\sigma} H \,.
\end{equation}
In particular, a term like $H^\dagger H \, \text{Tr} [F_{\mu\nu}^{(2)} F^{(2)\mu\nu}]$ would break the shift symmetry.  Expanding these expressions in the unitary gauge, we see that none of them involve a coupling of the Higgs to the photon.  At this order the $hZ\gamma$ coupling can then only be generated 
by the spurion couplings between the elementary and confining sectors and therefore will not be enhanced.

\section{Conclusion\label{sec:CO}}

In the unnatural, or split, composite Higgs model electroweak precision and flavour constraints are simply eliminated by requiring that $f\gtrsim 10$ TeV\@. This causes a splitting of the particle spectrum as the pseudo Nambu-Goldstone bosons are much lighter than the composite-sector resonances. In order to preserve gauge-coupling unification the model has a composite right-handed top quark and the strong sector must remain invariant under an $SU(5)$ global symmetry. This means that the low-energy spectrum generically contains the $SU(5)$ colour-triplet partner of the Higgs doublet, as well as a singlet scalar that plays the role of dark matter. In the minimal model residual symmetries related to proton and dark matter stability cause the colour-triplet scalar to decay via a dimension-six term in the Lagrangian and, since $f\gtrsim 10$ TeV, it can be metastable. Thus a long-lived colour-triplet scalar provides a distinctive experimental signal to test for unnaturalness.

R-hadron searches can be used to place limits on the colour-triplet mass and the current lower limit on a collider-stable ($c\tau \gtrsim 10$~m) colour-triplet from LHC Run-I results is around $845$ GeV\@. We have shown that with 300~fb$^{-1}$ of integrated luminosity at $\sqrt{s}=13$~TeV there is potential for a discovery up to a colour-triplet mass of 1.4 TeV or else it can be excluded up to 1.5 TeV. These limits significantly increase at a 100 TeV collider where, depending on the lifetime, triplets with masses ranging from 2 to 6 TeV can be discovered. Note that our limits from R-hadron searches are actually quite general, depending only on the mass and lifetime of the colour-triplet, and can be applied to any other model.  If the triplet decays in the inner detector (4~mm~$<r_{DV}<30$~cm) then displaced-vertex searches can be used to obtain limits. We find that the LHC can discover (exclude) colour-triplet masses up to 1.8~(1.9)~TeV for singlet masses below 450~GeV\@. At a 100~TeV collider the discovery reach is extended up to colour-triplet masses in the range 3-10~TeV depending on the singlet mass. There is also the possibility that the colour-triplet decays promptly when the mass $\gtrsim 4 $ TeV. In this case the colour-triplet can only be searched for at a future 100 TeV collider giving a potential exclusion for triplet masses ranging from 4 to 7 TeV, provided the singlet mass is less than around 900~GeV. Light colour-triplets can modify the Higgs coupling to gluons and current LHC limits lead to a lower bound on the mass $m_T\gtrsim100$~GeV\@. These limits can be improved upon at the HL-LHC or the ILC but remain weak compared to the direct detection limit of $m_S\gtrsim 150$ GeV from LUX, which implies that $m_T\gtrsim 300$ GeV assuming the singlet is the lightest stable particle.

Finally it should be noted that long-lived colour-triplet scalars are a sign of unnaturalness in composite Higgs models in much the same way that long-lived gluinos signal unnaturalness in split supersymmetric models. In both cases the experimental signals are quite similar because the decays produce jets and missing energy. Nevertheless there are differences related to the spin of the decaying particle and the particle(s) carrying the missing energy, as well as the large difference in the production cross-section. Given that current LHC results suggest that the Higgs potential may be tuned, it would therefore be worthwhile to study how these two unnatural possibilities could be distinguished at future colliders.

\section*{Acknowledgements}

We thank C.-P.~Yuan for useful discussions as well as Abi Soffer and Nimrod Taiblum for helpful correspondence regarding Ref.~\cite{1504.05162}. This work was supported by the Australian Research Council. TG was supported by the Department of Energy grant DE-SC0011842. AS was also supported by IBS under the project code IBS-R018-D1. PC is grateful to the Fine Theoretical Physics Institute for hospitality during the completion of this work.

\appendix
\newpage

\section{Four-Body Phase Space Integral\label{app:phase space}}

We present the calculation of the four-body phase space integral that is needed for obtaining the decay width of the colour-triplet scalar. We follow the common approach for many-body phase space integrals, and rewrite them as several two-body integrals. Given that the colour-triplet $T$ decays to $t^c b^c S S$, where $t(b)$ is the top (bottom) quark and $S$ is a singlet scalar, let $Q_1 = p_t + p_b$ and $Q_2 = p_{S_1} + p_{S_2}$. Note that the squared matrix element (\ref{eq:squaredM}) depends only on $Q_1^2$, and is independent of all other kinematic variables. The four-body phase space integral can be written
\begin{equation}
  \int d\Pi_4 (p_T ; p_t, p_b, p_{S_1}, p_{S_2}) = \int d\widetilde{\Pi}_2 (p_T; Q_1, Q_1) \, d\Pi_2 (Q_1; p_t, p_b) \, d\Pi_2 (Q_2; p_{S_1}, p_{S_2} ) \,,
\end{equation}
where
\begin{align}
  d\Pi_2 (p_a; p_1, p_2) & = \frac{d^4 p_1}{(2\pi)^4} \, \frac{d^4 p_2}{(2\pi)^4} \, 2\pi \theta(p_1^0) \delta (p_1^2 - m_1^2) \, 2\pi \theta(p_2^0) \delta (p_2^2 - m_2^2) \notag \\
  & \quad \times (2\pi)^4 \delta^{(4)} (p_a - p_1 - p_2) \,, \\
  d\widetilde{\Pi}_2 (p_a; p_1, p_2) & = \frac{d^4 p_1}{(2\pi)^4} \, \frac{d^4 p_2}{(2\pi)^4} (2\pi)^4 \delta^{(4)} (p_a - p_1 - p_2) \,.
\end{align}
We can then do the integrals over all momenta other than $Q_{1,2}$ trivially. Let us introduce the triangle function
\begin{equation}
  I(a, b) = 1 + a^2 + b^2 - 2a - 2b - 2ab \,.
  \label{eq:Idef}
\end{equation}
Then the two-body phase space integral may be written
\begin{equation}
  \int d\Pi_2 (p_a; p_1, p_2) = \frac{1}{8\pi} \biggl( \frac{2\lvert\vec{p}_1\rvert}{p_a^0} \biggr)_{COM} = \frac{1}{8\pi} \, \sqrt{I \biggl( \frac{m_1^2}{m_a^2} ,  \frac{m_2^2}{m_a^2} \biggr)}  \,.
\end{equation}
The first result is the well-known expression for the two-body phase space in the centre of mass frame; the second result expresses this in Lorentz-invariant form. Since the integral is manifestly Lorentz-invariant this result holds in all frames. In the two specific cases we require this simplifies further. Neglecting the bottom quark mass we have
\begin{align}
  \int d\Pi_2 (Q_1; p_t, p_b) & = \frac{1}{8\pi} \, \biggl( 1 - \frac{m_t^2}{Q_1^2} \biggr) \,, \\
  \int d\Pi_2 (Q_2; p_{S_1}, p_{S_2}) & = \frac{1}{16\pi} \, \sqrt{ 1 - \frac{4 m_S^2}{Q_2^2}} \,.
\end{align}
The additional factor of one-half in the latter equation is due to the presence of identical final states.

Next, we rewrite the integral over $Q_1$ and $Q_2$. It is easy to see that, if $p_{1,2}^0$ are constrained positive,
\begin{equation}
  d\widetilde{\Pi}_2 (p_a; p_1, p_2) = \frac{dm_1^2}{2\pi} \, \frac{dm_2^2}{2\pi} \, d\Pi_2 (p_a; p_1, p_2) \,.
\end{equation}
This condition applies to $Q_{1,2}$. Therefore we may write
\begin{equation}
  \int d\widetilde{\Pi}_2 (p_t; Q_1, Q_1) = \int \frac{dQ_1^2}{2\pi} \, \frac{dQ_2^2}{2\pi} \, \frac{1}{8\pi} \sqrt{I \biggl( \frac{Q_1^2}{m_T^2} ,  \frac{Q_2^2}{m_T^2} \biggr)} \,.
\end{equation}
Putting all of this together, we have the final result
\begin{equation}
  \int d\Pi_4 (p_T ; p_t, p_b, p_{S_1}, p_{S_2}) = \frac{1}{2^{12}\pi^5} \int dQ_1^2 \, dQ_2^2 \sqrt{I \biggl( \frac{Q_1^2}{m_T^2} ,  \frac{Q_2^2}{m_T^2} \biggr)} \, \biggl( 1 - \frac{m_t^2}{Q_1^2} \biggr) \sqrt{ 1 - \frac{4 m_S^2}{Q_2^2}} \,.
\end{equation}
Finally we need the limits on the integral. It is straightforward to see that the absolute bounds on $Q_1^2$ are
\begin{equation}
  m_t^2 < Q_1^2 < (m_T - 2m_S)^2 \,.
\label{eq:Q1bounds}\end{equation}
The lower bound occurs when the $b$ quark is produced at rest, and the upper bound when the two $S$ are at rest. For any given $Q_1^2$ there is an upper bound on $Q_2^2$ and so
\begin{equation}
  4m_S^2 < Q_2^2 < \left(m_T - \sqrt{Q_1^2}\right)^2 \,.
\label{eq:Q2bounds}\end{equation}
The lower bound arises from when the two $S$ are at rest, while the upper bound is obtained when they are back-to-back.

\section{Displaced-Vertex Search Validation} \label{app:DV_validation}

Given the challenges involved in recasting displaced searches and the various assumptions that must be made, it is important to check the validity of our implementation against the full experimental analysis. We have therefore also simulated events for one of the signal processes considered in the ATLAS paper~\cite{1504.05162}. We have chosen the case of a long-lived gluino decaying to two top quarks and a 100 GeV neutralino since this most closely resembles the final-state that is produced by the decay of our colour-triplet. 

\begin{figure}[H]
  \centering
  \begin{minipage}[b]{0.49\textwidth}
    \centering
    \includegraphics[width=0.9\textwidth]{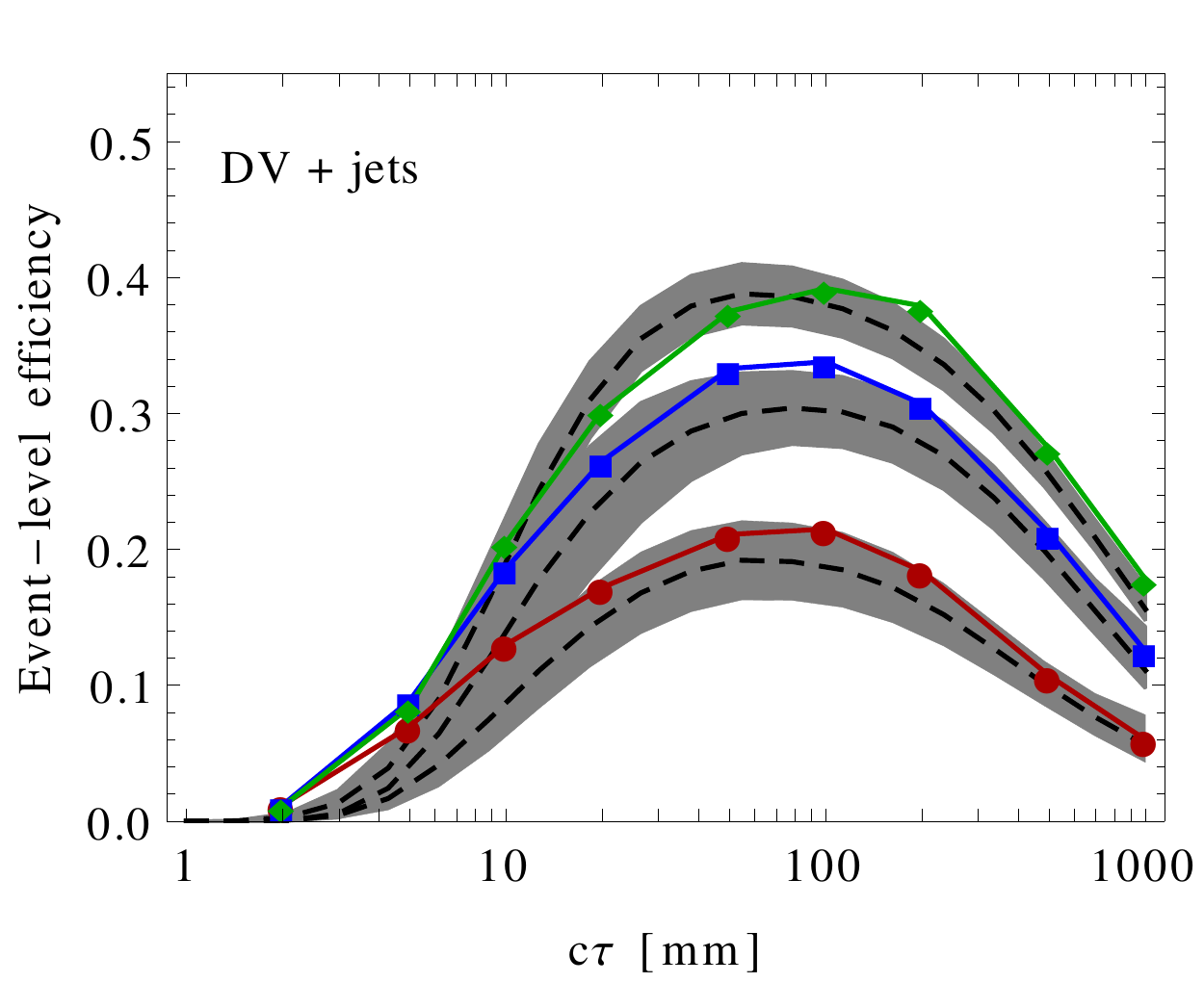}
  \end{minipage}
  \begin{minipage}[b]{0.49\textwidth}
    \centering
    \includegraphics[width=0.9\textwidth]{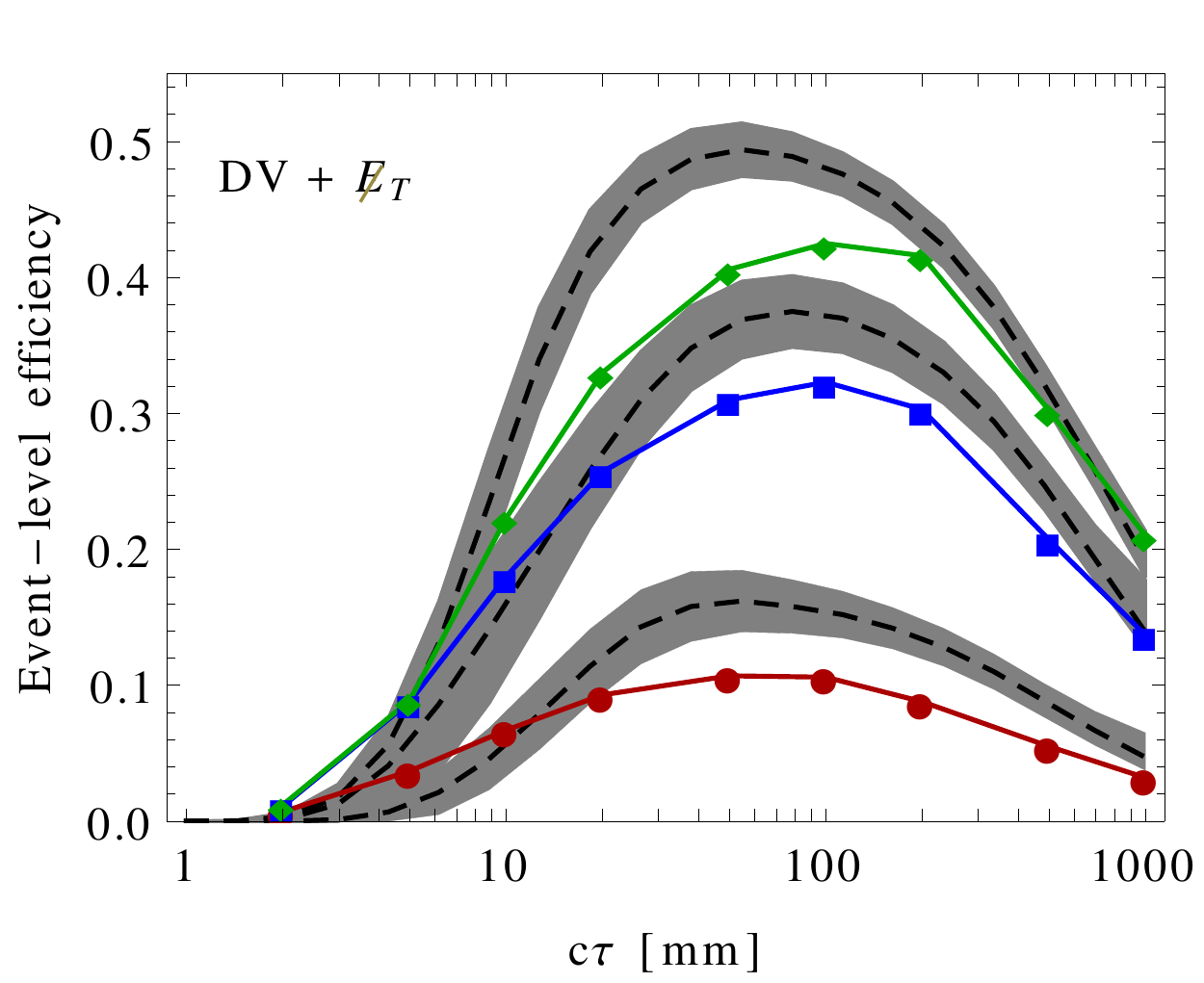}
  \end{minipage}
  \caption{Comparison of the event-level efficiencies from our analysis (data points) and the ATLAS analysis (shaded regions) for the case of a long-lived gluino decaying to $tt\tilde\chi_0$. From top to bottom the curves correspond to gluino masses of 1400, 1000 and 600 GeV. The left and right panels are for the DV+jets and DV+$\slashed{E}_T$ channels respectively.} \label{fig:DV_validation}
\end{figure}

In Figure~\ref{fig:DV_validation} we compare the event-level efficiencies obtained from our analysis (data points) with the results reported by ATLAS (shaded regions) for both the DV+jets and DV+$\slashed{E}_T$ channels. Overall we find that our analysis gives reasonably good agreement with the full experimental analysis, especially in the DV+jets channel. The discrepancies in the DV+$\slashed{E}_T$ channel suggest that our assumptions regarding the reconstruction of the decay products from displaced R-hadron decays leads to an underestimate of the missing energy. The difference in signal efficiency is not expected to have a significant effect on the exclusion limits we derive, especially at higher center-of-mass energies where the expected missing energy from our signal can be significantly greater than the experimental cuts.

\bibliographystyle{biblio_style}
\bibliography{LongLivedTriplet}

\end{document}